\documentclass[%
reprint,
superscriptaddress,
 amsmath,amssymb,
 aps,
]{revtex4-2}

\usepackage{graphicx}
\usepackage{hyperref}
\usepackage{dcolumn}
\usepackage{bm}
\usepackage{braket}
\usepackage{blindtext}
\usepackage{ragged2e} 
\usepackage{lipsum}

\usepackage{siunitx}
\DeclareSIUnit{\torr}{Torr}
\DeclareSIUnit{\mV}{mV}
\DeclareSIUnit{\mT}{mT}
\DeclareSIUnit{\nm}{nm}
\DeclareSIUnit{\rpm}{rpm}
\DeclareSIUnit{\sccm}{sccm}
\DeclareSIUnit{\angstrom}{\text{\AA}}
\DeclareSIUnit{\mK}{mK}
\DeclareSIUnit{\mum}{\micro\meter}

\usepackage{xcolor}
\definecolor{emily}{HTML}{5F00BA}
\definecolor{mario}{HTML}{5250FF}

\newcommand{\gateratio}{\Delta V/\Delta V_{1}}
\newcommand{\fieldratio}{\Delta B_{1}/\Delta B}

\DeclareUnicodeCharacter{2212}{-}

\begin{document}

\preprint{APS/123-QED}

\title{Trapping $e/4$ quasiparticles in bilayer graphene}

\author{Mario Di Luca}
\thanks{These authors contributed equally.}
\author{Emily Hajigeorgiou}
\thanks{These authors contributed equally.}
\author{Ning Ma}
\author{Alexandra Waldherr}
\affiliation{Institute of Physics, École Polytechnique Fédérale de Lausanne (EPFL), CH-1015 Lausanne, Switzerland}

\author{Kenji Watanabe}

\author{Takashi Taniguchi}
\affiliation{International Center for Materials Nanoarchitectonics,
National Institute for Materials Science, 1-1 Namiki, Tsukuba 305-0044, Japan}

\author{Mitali Banerjee}
\email{mitali.banerjee@epfl.ch}
\affiliation{Institute of Physics, École Polytechnique Fédérale de Lausanne (EPFL), CH-1015 Lausanne, Switzerland}
\affiliation{Center for Quantum Science and Engineering (QSE Center), École Polytechnique Fédérale de Lausanne (EPFL), CH-1015 Lausanne, Switzerland}

\begin{abstract}
Measuring the charge of the quasiparticles hosted by even-denominator fractional quantum Hall (FQH) states is essential to identify the topology of their ground state. Here, we use a gate-defined antidot in bilayer graphene, with an additional gate to control only the antidot potential, to measure the charge of the quasiparticles trapped around it in even-denominator FQH states. We observe a localized charge of $e/4$ at $\nu=-5/2$, $-1/2$, and $3/2$, consistent with the minimal excitation expected for leading candidate even-denominator ground states, and $e/3$ at the hole-conjugate state $\nu=2/3$. We further show that increasing the coupling between the antidot-bound states and extended edge states drives a crossover between two regimes, characterized by the minimal-excitation gate-voltage period and approximately twice that period, respectively. We discuss two possible explanations for this crossover: quasiparticle bunching and a crossover between distinct antidot transport regimes. Our results, together with previous observations of the daughter states, show that the even-denominator FQH states in bilayer graphene are compatible with a non-Abelian ground state, and that their quasiparticles can be localized around a quantum Hall antidot, a necessary ingredient for topological quantum computation.
\end{abstract}

\maketitle

\section*{Introduction}

Non-Abelian quasiparticles are among the most sought-after excitations in condensed-matter physics. Their braiding transforms the many-body degenerate ground state in a way that depends only on the topology of the exchange path, making them promising building blocks for fault-tolerant quantum computation~\cite{Kitaev2003Jan, DasSarma2005Apr, Nayak2008Sep}. Even-denominator fractional quantum Hall (FQH) states are leading candidates for hosting non-Abelian excitations. Unlike the odd-denominator FQH states of the conventional hierarchy, they occur at half-filled Landau levels, where the standard composite-fermion picture predicts a compressible Fermi sea rather than an incompressible state~\cite{Jain1989Jul, Jain2007Mar}. Their observation therefore requires a different mechanism, in which composite fermions pair to form an incompressible quantum Hall fluid~\cite{Greiter1991Jun, Moore1991Aug}. A central prediction of these paired states is the emergence of quasiparticles carrying charge $e/4$~\cite{Bonderson2006Jan, Levin2007Dec, Ma2019Jul, Feldman2021Jun}. Such excitations are expected for both Abelian and non-Abelian candidate states, thus their observation is a necessary but not sufficient condition for non-Abelian excitations. Demonstrating that such quasiparticles can be experimentally detected and localized is therefore a key step toward controlling the elementary excitations of even-denominator FQH states and ultimately probing their topological properties.

The first evidence for charge-$e/4$ quasiparticles was observed in GaAs $\nu=5/2$ systems, where shot-noise~\cite{Dolev2008Apr}, local-compressibility~\cite{Venkatachalam2011Jan}, and quasiparticle-tunneling~\cite{Radu2008May, Lin2012Apr, Baer2014Aug, Fu2016Nov} measurements provided evidence for quarter-charge excitations. More recently, Bernal-stacked bilayer graphene (BLG) has emerged as a particularly promising platform for studying even-denominator FQH states, owing to the observation of multiple such states~\cite{Li2017Oct, Huang2022Jul, Assouline2024Jan, Kumar2025Aug} and the high degree of tunability offered by its internal degrees of freedom. This tunability provides access to distinct ground states, such as candidate Pfaffian and anti-Pfaffian orders, making BLG an attractive platform for probing their quasiparticle properties. 

Fabry--P\'erot interferometry in BLG has provided phase-sensitive probes of these states~\cite{Kim2026Jan, Henzinger2026Mar, Kim2026Mar}. In recent interferometry experiments, the observed periodicity is consistent with charge-$e/2$ interfering quasiparticles, rather than the expected charge-$e/4$ quasiparticles, while the localized excitations enclosed by the interferometer carry charge $e/4$. This motivates a complementary approach to probe the elementary quasiparticle charge of these states, through controlled localization. A quantum Hall antidot offers an alternative route in which quasiparticles, supported by the surrounding FQH fluid, can be localized around a potential hill, and the charge of the localized quasiparticles is directly obtained from conductance oscillations~\cite{Goldman1995Feb, Franklin1996Jul, Sim2008Feb, Kou2012Jun, Mills2020Nov, DiLuca2026Aug}.

Here, we use a gate-defined antidot in BLG to perform charge spectroscopy of even-denominator FQH states. At $\nu=-5/2$, $-1/2$, and $3/2$, we observe Coulomb-dominated (CD) oscillations corresponding to the addition of charge $e/4$, demonstrating that $e/4$ quasiparticles can be localized and accessed in these states. Additionally, we observe oscillations at the hole-conjugate FQH state $\nu=2/3$, corresponding to the addition of charge $e/3$ to the antidot. Furthermore, by tuning the coupling between the antidot-bound states and the extended edge states, we observe a crossover from the minimal-excitation gate-voltage period to approximately twice that value. We discuss two possible interpretations of this crossover: charge-$e/2$ addition, for example through bunching of two charge-$e/4$ quasiparticles, or a crossover toward an Aharonov--Bohm (AB) dominated or mixed transport regime in which the gate-voltage periodicity is no longer set solely by Coulomb charge addition. The observation of a similar crossover at the Abelian hole-conjugate state $\nu=2/3$ shows that doubled periodicity is not unique to even-denominator states. However, the absence of doubling in the conventional odd-denominator states studied here suggests that the effect is linked to the specific edge-structure of the underlying state, rather than being an inherent feature of the device geometry.

\section*{Quantum Hall antidot device}

\begin{figure*}[tp!]
 \includegraphics[width = 0.95\textwidth]{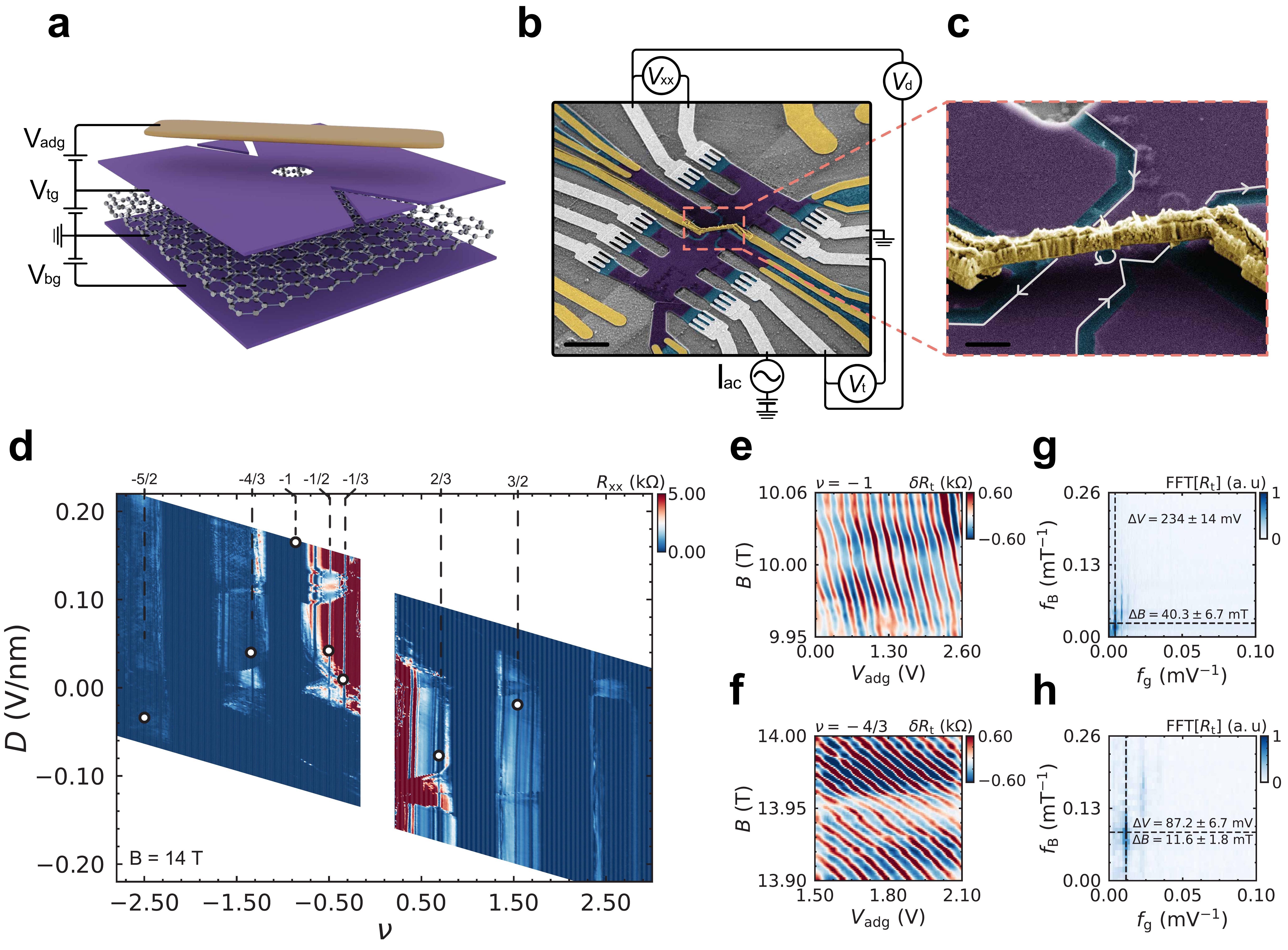}
 \begin{center}
 \caption{\textbf{Gate-defined antidot in bilayer graphene controlled by an air-bridge.} \textbf{a}, Schematic of the gate-defined antidot in bilayer graphene controlled by an air-bridge. The top and bottom graphite gates are shown in purple, while the air-bridge is shown in yellow. The top and bottom hBN are not shown for clarity. \textbf{b}, False-color SEM of a device with the air-bridge. The top graphite gate is shown in purple, the exposed hBN regions, where the top graphite has been etched away, are shown in blue, and the metallic gates, including the suspended air-bridge gate, are shown in yellow. We apply an AC current $I_\mathrm{ac} = \qty{0.2-1}{nA}$ and measure the voltage drop across the AD longitudinally ($V_\mathrm{t}$) and diagonally ($V_\mathrm{d}$) together with the voltage drop on its side ($V_\mathrm{xx}$). The current is drained through a CG, a contact shorted to the cold finger. The scale bar is \qty{2}{\micro\meter}. \textbf{c}, A zoomed false-color SEM showing the air-bridge suspended over the etched hole in the top graphite. The ideal path of the edge states at $\nu = 1$ is shown in white. The scale bar is \qty{400}{\nm}. \textbf{d}, Longitudinal resistance $R_\mathrm{xx}$ measured outside the AD as a function of the filling factor $\nu$ and the displacement field $D$ at $B = \qty{14}{\tesla}$. The white circles indicate the regions where we have measured the oscillations. \textbf{e-f}, Variation of the transmitted resistance oscillations as a function of the ADG voltage, $V_\mathrm{adg}$, and the magnetic field $B$ at $\nu = -1$ and $-4/3$, respectively. \textbf{g-h}, 2D-FFT of the respective oscillations.} 
 \label{fig:figure_1}
 \end{center}
\end{figure*}

Quantum Hall antidots (AD), or simply antidots, have previously been used to trap and measure the charge of trapped quasiparticles in several odd-denominator FQH states in GaAs~\cite{Goldman1995Feb, Franklin1996Jul, Kou2012Jun}, monolayer graphene~\cite{Mills2020Nov}, and BLG~\cite{DiLuca2026Aug}. Here, we extend this approach to even-denominator FQH states in Bernal-stacked bilayer graphene.

Compared with the previous device design \cite{DiLuca2026Aug}, we introduce a suspended air-bridge gate that predominantly couples to the AD region, which we refer to as the antidot gate (ADG). This local control is essential for resolving oscillations in even-denominator FQH states, whose incompressible plateaus are substantially narrower than the odd-denominator FQH states. Using the top gate alone would rapidly tune the bulk filling outside the incompressible plateau, making it impractical to resolve AD oscillations over the narrow stability range of these states. A schematic of the device is shown in Fig.~\ref{fig:figure_1}a. The top and bottom graphite gates are shown in purple, while the suspended air-bridge is shown in yellow. The top and bottom hBN layers are omitted for clarity. A false-color scanning electron micrograph (SEM) of the device is shown in Fig.~\ref{fig:figure_1}b, with a zoom-in of the AD region shown in Fig.~\ref{fig:figure_1}c. The top graphite gate is shown in purple, the exposed hBN regions, where the top graphite has been etched away, are shown in blue, and the metallic gates, including the suspended air-bridge gate, are shown in yellow.

Even-denominator FQH states have small activation gaps, on the order of \SI{1}{\kelvin} at the magnetic field of $B = \SI{14}{\tesla}$ at which we carry out our measurements~\cite{Assouline2024Jan, Kumar2025Aug}, making low electron temperature particularly important. To improve thermalization, we use a cold-grounded contact (CG), that is, a contact directly shorted to the cold finger. However, this configuration prevents us from directly measuring the transmitted current. Therefore, except for DC-bias characterization, all measurements are performed in a current-biased configuration, where we source an AC current $I_\mathrm{ac}=\qty{0.2-1}{nA}$ and measure the voltage drop across the AD, $V_\mathrm{t}$, the diagonal voltage drop, $V_{\mathrm{d}}$, and the longitudinal voltage drop on the side of the AD, $V_\mathrm{xx}$. The measurement schematic is shown in Fig.~\ref{fig:figure_1}b.

In the following, we report oscillations in the variation of transmitted resistance $\delta R_{\mathrm{t}}$, to facilitate comparison between filling factors. The diagonal conductance is measured simultaneously to verify that the filling factor of the region between the AD-bound states and the extended edge states, hereafter called the constriction filling factor, remains close to that of the bulk. In parallel, the longitudinal resistance measured outside the AD confirms that the surrounding bulk stays incompressible throughout the characterization.

We start by characterizing the device at a magnetic field of $B = 14$\,T. Figure~\ref{fig:figure_1}d shows the longitudinal resistance $R_\mathrm{xx}$, measured on the side of the AD, as a function of the bulk filling factor $\nu$ and the displacement field $D$. At this magnetic field, the eightfold degeneracy of the two lowest Landau levels of BLG is fully lifted, and the levels are fully polarized in spin, valley, and orbital index. In particular, for $|\nu| \leq 4$, and low displacement field, we observe only states with the same spin but different orbital, $N = 0, 1$, and valley, $K$ and $K'$, quantum numbers~\cite{Hunt2017Oct, Li2018Jan}, with the even-denominator FQH states appearing in the $N = 1$ region. We observe well developed even-denominator FQH states at $\nu = -5/2$, $-1/2$ and $+3/2$, and developing states at $\nu = +1/2$ and $+5/2$.

\section*{Oscillations at integer and odd-denominator states}

We tune the AD following the procedure described in Ref.~\cite{DiLuca2026Aug}. The bottom gate (BG) defines the AD potential, the top gate (TG) sets the bulk filling factor around the AD to a FQH state, while the ADG drives the oscillations. Sweeping $V_\mathrm{adg}$ raises or lowers the local potential hill, changing the area enclosed by an orbit of given energy. Since the flux threading each AD-bound state must stay quantized in units of $\phi_0 = h/e$, the enclosed area is fixed at a given field, and the bound states shift in energy along the AD potential. Resonant tunneling through the AD occurs whenever one of these levels aligns with the electrochemical potential of the extended edge states. Importantly, the gate periodicity measured here characterizes the discrete change in the localized charge bound to the AD between successive resonances, which is not necessarily equal to the charge of the individual tunnelling quasiparticles.

The localized quasiparticle charge $q$ can be extracted using two complementary methods, both applicable in the CD regime~\cite{DiLuca2026Aug}. The first uses the slope of the oscillations in the gate-magnetic field plane~\cite{Goldman1995Feb, Franklin1996Jul}. This method has two limitations: the charge it returns cannot be uniquely distinguished from the filling factor at which the oscillations are measured~\cite{DiLuca2026Aug}, and it assumes that the gate varies only the density around the AD. In our geometry the ADG mainly controls the AD potential, so this approach is not applicable here (see Supp. Material). The second method uses the ratio between the gate voltage period at the filling factor of interest $\Delta V$ and the period corresponding to the addition of a single electron, $\Delta V_1$, taken as a reference. We therefore extract the quasiparticle charge from the normalized gate voltage period, $q/e = \gateratio$. Similarly, we use the ratio between the magnetic field period corresponding to one flux quantum $\phi_0$, denoted $\Delta B_1$ and taken as a reference, and the magnetic field period $\Delta B$ measured at a given filling factor, to extract the number of quasiparticles transferred per flux quantum, $p = \fieldratio$. Throughout this work we use $\nu = -1$ at $B = 10$\,T as the reference state (see Supp. Material).

\begin{figure*}[thp!]
 \includegraphics[width = 0.95\textwidth]{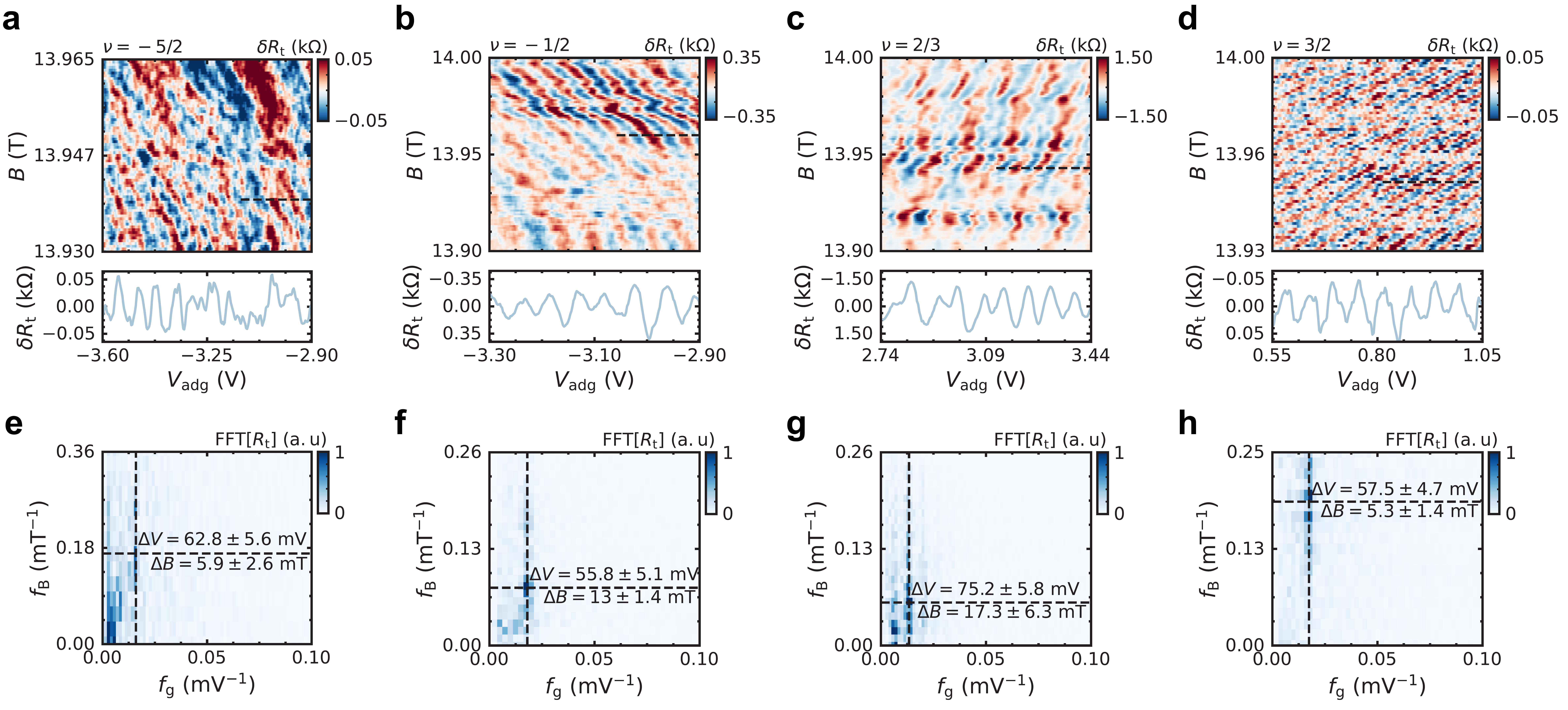}
 \begin{center}
 \caption{\textbf{Minimal-excitation oscillations at even-denominator and hole-conjugate FHQ states.} \textbf{a--d}, Variation of the transmitted resistance oscillations $\delta R_\mathrm{t}$ as a function of the ADG voltage $V_\mathrm{adg}$ and the magnetic field $B$ for filling factors $\nu = -5/2, -1/2, 2/3$ and $3/2$ respectively. For each panel, the lower light blue trace shows a line cut through the oscillations along the black dashed line indicated in the corresponding map. \textbf{e--h}, 2D-FFT of the oscillations with the corresponding extracted ADG ($\Delta V$) and magnetic field ($\Delta B$) periods for each filling factor.} 
 \label{fig:figure_2}
 \end{center}
\end{figure*}

Figures~\ref{fig:figure_1}e--f show $\delta R_\mathrm{t}$ as a function of the ADG voltage, $V_\mathrm{adg}$ and the magnetic field $B$, at $\nu = -1$ and $-4/3$. Since these two states have already been characterized in detail using the TG~\cite{DiLuca2026Aug}, we use them here to calibrate the ADG oscillations and verify the analysis procedure, see the Supp. Material for the characterization using the TG. From the two-dimensional fast Fourier transform (2D-FFT) in Fig.~\ref{fig:figure_1}g--h, we find that the ratio between the ADG voltage period at $\nu = -4/3$ and that at $\nu = -1$ is equal to $\gateratio = 0.37 \pm 0.04$, in agreement with the value $q/e = 1/3$ expected for CD antidot oscillations previously reported. The ratio between the magnetic field period at $\nu = -1$ and that at $\nu = -4/3$ is equal to $\fieldratio = 3.47 \pm 0.79$, consistent with the expected $p = 4$ quasiparticles per flux quantum, within the uncertainty. In addition, $\Delta B_1$ gives access to the antidot diameter, which we find to be $D = \SI{361\pm 30}{\nm}$, consistent with the lithographic diameter of approximately \SI{210}{\nm} and a top hBN thickness of approximately \SI{46}{\nm}, see Supp. Material. 

We therefore conclude that the ADG provides reliable local control of the AD and that our analysis reproduces the previously benchmarked behaviour in a similar device~\cite{DiLuca2026Aug}. In the following, we use the ADG to drive the oscillations in all the filling factors studied here.

\section*{Oscillations at even-denominator and hole-conjugate states}

Next, we study the AD oscillations as a function of $V_{\mathrm{adg}}$ and magnetic field $B$ at even-denominator FQH states $\nu=-5/2, -1/2,~3/2$, and at the hole-conjugate FQH state $\nu = 2/3$. The $\delta R_\mathrm{t}$ oscillations are shown in Fig.~\ref{fig:figure_2}a--d for these states, including a line cut along the black dashed line. Each state is accompanied by the corresponding 2D-FFT (Fig.~\ref{fig:figure_2}e--h) used to extract the gate-voltage and magnetic-field periods (see Supp. Material and Methods for details on the analysis).

For the even-denominator states at $\nu = -5/2$, $-1/2$ and $3/2$ (Fig.~\ref{fig:figure_2}a,b,d) we extract ADG voltage periods in the range \SIrange{56}{63}{\mV}, giving respective normalized periods of $\gateratio = 0.27\pm 0.03 $ for $\nu = -5/2$, $\gateratio = 0.24\pm 0.03$ for $\nu = -1/2$ and $\gateratio = 0.25\pm 0.03$ for $\nu = 3/2$. Within the uncertainty, all three filling factors yield the same gate-voltage period ratio, $\gateratio \approx 0.25$, consistent with the addition of a localized quasiparticle with charge $e/4$ derived from $\gateratio = q/e$. The corresponding magnetic-field periods, $\Delta B_{-5/2}=\SI{5.9 \pm 2.6}{\mT}$, $\Delta B_{-1/2}=\SI{13 \pm 1.4}{\mT}$, and $\Delta B_{3/2}=\SI{5.3 \pm 1.4}{\mT}$, are consistent with measurements in the CD regime. Within uncertainty, these periods correspond approximately to $p=10$, $2$, and $6$ quasiparticles transferred per flux quantum, respectively~\cite{Read1996Dec, DasSarma2005Apr}. We therefore conclude that, for all even-denominator states studied here, the observed oscillations are consistent with sequential changes in the localized AD charge in units of the minimal excitation charge, $q=e/4$.

\begin{figure*}[tp!]
 \includegraphics[width = 0.95\textwidth]{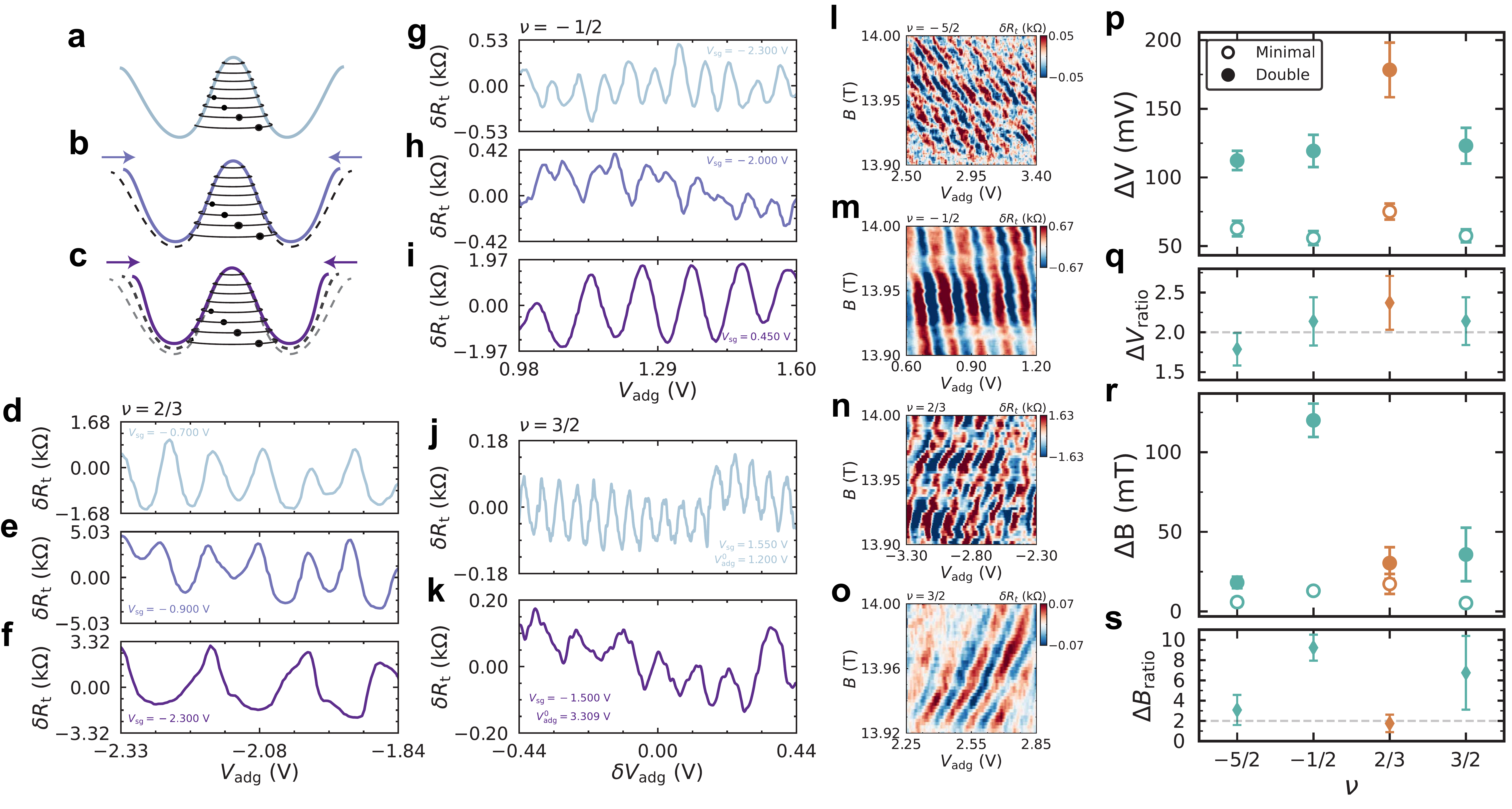}
 \begin{center}
\caption{\textbf{Crossover from minimal excitation to double-gate-period as a function of side gate voltage $V_{sg}$.} \textbf{a--c}, Schematic of the AD potential landscape with its quantized bound states. Making the side gates increasingly positive (negative) on the hole (electron) side brings the extended edge states closer to the AD-bound states, increasing their coupling and resulting in a doubling of the measured ADG voltage period. \textbf{d--f}, Transmitted resistance variation $\delta R_\mathrm{t}$ oscillations at $\nu=-1/2$ as a function of AD-gate voltage $V_{\mathrm{adg}}$ for increasingly positive side-gate voltage. The oscillation period evolves from  $\Delta V \approx0.25\,\Delta V_{1}$ in the weak-coupling regime (\textbf{d}) to $\Delta V\approx 0.5\,\Delta V_{1}$ in the stronger-coupling regime (\textbf{f}), with an intermediate regime where both periodicities are superimposed (\textbf{e}). \textbf{g--i}, Period doubling at $\nu=2/3$ as the side gates are made increasingly negative. \textbf{j--k}, Period doubling at $\nu=3/2$ as the side gates are made increasingly negative. \textbf{l--o}, Transmitted resistance oscillations $\delta R_\mathrm{t}$ as a function of the ADG voltage and the magnetic field $B$ in the double-gate-period region for $\nu = -5/2, -1/2, 2/3$ and $3/2$. \textbf{p}, ADG voltage period $\Delta V$, and \textbf{q}, the ratio between the double regime and minimal excitation gate periods ($\Delta
V_{\mathrm{double}}/\Delta V_{\mathrm{minimal}}$) as a function of the filling factor. \textbf{r}, Magnetic field period $\Delta B$, and \textbf{s}, the ratio between the double regime and minimal excitation magnetic field periods ($\Delta
B_{\mathrm{double}}/\Delta B_{\mathrm{minimal}}$) as a function of the filling factor. Open circles denote the minimal excitation and filled circles the double regime periods. In addition, two colours are used: turquoise for the even-denominator FQH states and ochre for the hole-conjugate FQH states.}
 \label{fig:figure_3}
 \end{center}
\end{figure*}

At $\nu = 2/3$ (Fig.~\ref{fig:figure_2}c) we find a gate voltage period ratio of $\gateratio = 0.32 \pm 0.03$ and a magnetic field period ratio of $\fieldratio = 2.32 \pm 0.93$, compatible with $q = e/3$ and $p = 2$. This data set shows substantial gate instability, therefore we validate the periodicity with a second, independent method following Ref.~\cite{Werkmeister2025Apr}: we build a histogram of the oscillations by repeatedly acquiring the same trace, and repeat the procedure at several magnetic fields. Tracking the shift of the dominant histogram peak with magnetic field gives an estimate of $\Delta B$ that agrees with the value extracted from the 2D-FFT (see Supp. Material). Our measurement at $\nu=2/3$ therefore indicates sequential changes in the localized AD charge in units of $e/3$. We attribute the difference with respect to previous reports, where the dominant charge was $2e/3$~\cite{Kou2012Jun, DiLuca2026Aug}, to a lower electron temperature (of the order of 30--50\,mK compared to the 150\,mK in Ref.~\cite{DiLuca2026Aug}), and to the improved control over the coupling between the AD-bound states and the extended edge states offered by our ADG geometry.

\section*{Beyond the weak-backscattering regime}

In the previous sections, we have shown that the expected minimal excitation can be clearly resolved across odd-denominator, even-denominator, and hole-conjugate FQH states, provided that the AD is weakly coupled to the extended edge states and the fractional bulk in the constriction remains only weakly perturbed. This indicates that the absence of resolved charge-$e/4$ and $e/3$ periodicities in earlier measurements of even-denominator FQH states~\cite{Kim2026Jan, Henzinger2026Mar, Kim2026Mar} and the hole-conjugate state $\nu=2/3$~\cite{Kou2012Jun, DiLuca2026Aug} is unlikely to be intrinsic to the states themselves. Among the possible experimental factors, such as electron temperature and smoothness of the confining potential, the coupling between the AD-bound states and the extended edge states can be tuned in situ in our device. In this section, we therefore study how the minimal-excitation oscillations evolve as this coupling is increased. In practice, we control the coupling with the side gates: increasingly negative (positive) voltages for electron (hole) states bring the extended edges closer to the AD, as illustrated in Fig.~\ref{fig:figure_3}a--c.

Figures~\ref{fig:figure_3}d,g,j show the transmitted resistance oscillations as a function of the ADG voltage at $\nu = -1/2$, $2/3$ and $3/2$ respectively, measured in the same minimal excitation regime. Starting from this configuration, we then repeat the measurement at a series of side-gate voltages $V_\mathrm{sg}$, keeping the TG and BG voltages fixed and, where the oscillations remain visible, the same ADG range.

As the absolute value of the constriction filling factor is reduced, the minimal-excitation oscillations evolve into an intermediate regime in which a doubled-period component is superimposed on the minimal-excitation period at $\nu=2/3$ and $-1/2$, as shown in Fig.~\ref{fig:figure_3}e,h. Upon further increasing the coupling, the doubled period becomes the dominant contribution, and the minimal-excitation periodicity is no longer resolved, resulting in an effective doubling of the ADG voltage period. We note that for $\nu = 3/2$ we could not find a region of intermediate transition where both periodicities are present. 

Figures~\ref{fig:figure_3}l--o report $\delta R_\mathrm{t}$ oscillations as a function of the ADG voltage and the magnetic field in the double-gate-period region for the filling factors $\nu = -5/2, -1/2, 2/3$ and $3/2$. Also in this case, the period values obtained at $\nu = 2/3$ have been cross-checked with the histogram method reported in the Supp. Material.

The extracted ADG and magnetic field periods are reported in Fig.~\ref{fig:figure_3}p and Fig.~\ref{fig:figure_3}r, respectively, where open circles denote the minimal excitation oscillations and filled circles the double-gate-period ones. Figure~\ref{fig:figure_3}q shows the ratio between the gate voltage period of the double-gate-period oscillations and that of the minimal excitation: for all filling factors this ratio falls close to $2$. The behavior of the magnetic-field ratio is less clear-cut. Only $\nu=-5/2$ and $\nu=2/3$ are consistent with a ratio of $2$, while the remaining filling factors lie above this value; $\nu=3/2$ approaches close to $2$ within uncertainty, while an extremely large magnetic field periodicity occurs for $\nu=-1/2$.

\section*{Discussion}

\begin{figure}[tph!]
 \includegraphics[width = 0.465\textwidth]{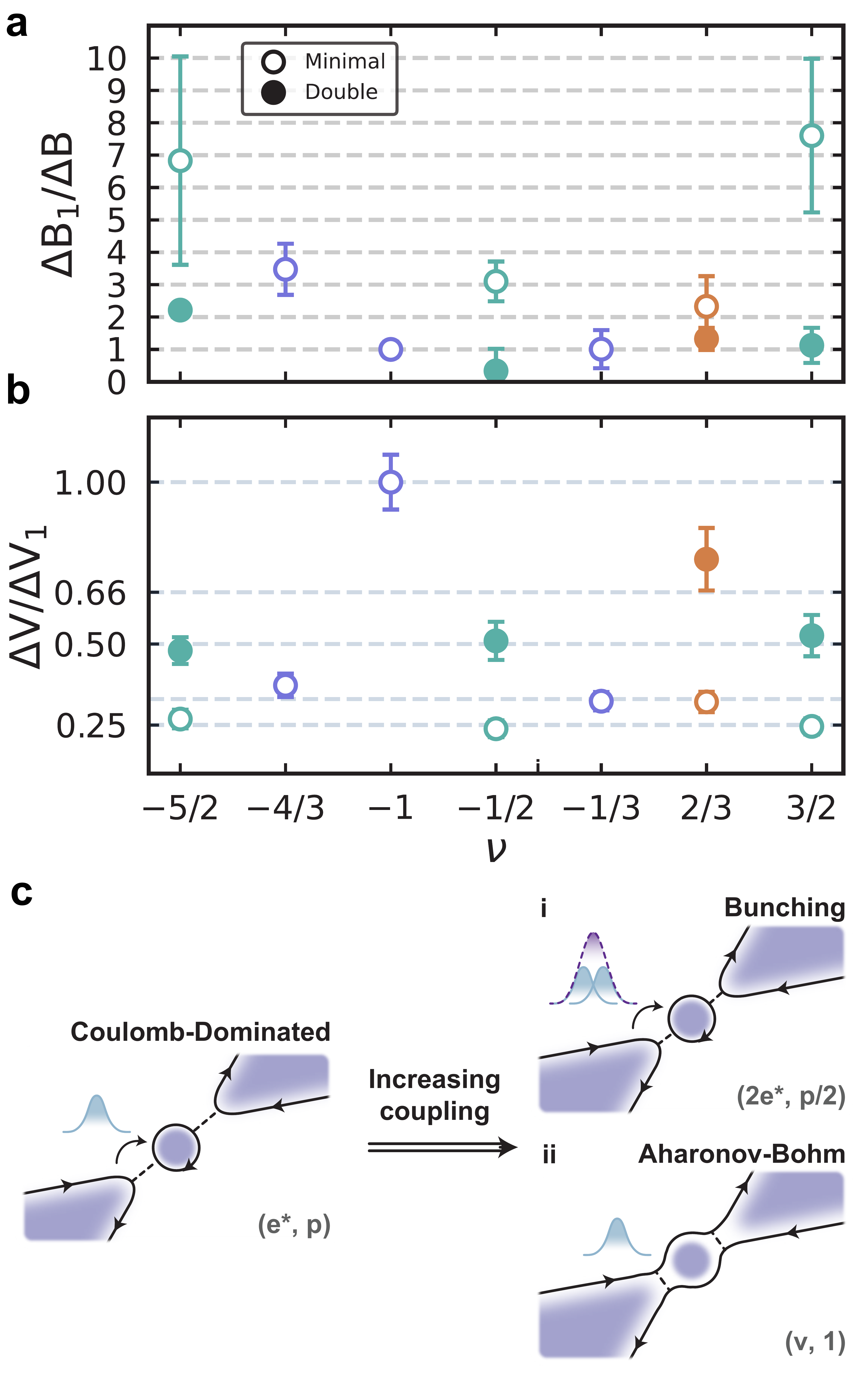}
 \begin{center}
 \caption{\textbf{Normalized gate voltage and magnetic field ratios.} \textbf{a}, Ratio of the magnetic field at filling factor $\nu = -1$, $\Delta B_1$ used as a reference and corresponding to a flux periodicity of $\phi_0$, to the magnetic field period at a given filling factor $\Delta B$ as a function of the filling factor $\nu$. \textbf{b}, Ratio of the ADG voltage period at a given filling factor $\Delta V$ to the ADG voltage period at $\nu = -1$, used as reference $\Delta V_1$, and corresponding to the addition of one electron to the AD. Open circles denote the minimal excitation and filled circles the double periodicity. In addition, three colours are used: light purple for the electron-like odd-denominator fractional and integer quantum Hall states, turquoise for the even-denominator FQH states, and ochre for the hole-conjugate FQH states. \textbf{c}, Schematic of two possible interpretations for the observed gate periodicity doubling. In the weak-backscattering regime, individual quasiparticles can tunnel between the extended edge states and the AD-bound states, producing discrete changes in the localized antidot charge. As the coupling of the AD-bound states to the extended edge states is increased, two possible scenarios unfold: i) it becomes energetically favorable for two quasiparticles to tunnel together, a process generally referred to as bunching; or ii) the system crosses over toward an AB regime, in which the gate-voltage period is no longer directly determined by the quasiparticle charge. The gray labels on the bottom right in each inset report the expected value of the $\gateratio$ and $\fieldratio$, respectively.} 
 \label{fig:figure_4}
 \end{center}
\end{figure}

Figures~\ref{fig:figure_4}a,b summarize the normalized period ratios in magnetic field, $\fieldratio$, and in gate voltage, $\gateratio$, for all the filling factors studied. The states are labeled by three colors: electron-like states, comprising the integer and the odd-denominator fractional states (light purple); even-denominator states (turquoise); and hole-conjugate states (ochre).

When the AD is weakly coupled to the extended edge states, the extracted charge values agree with the expected minimal excitation: $e/4$ at the even-denominator filling factors $\nu = -5/2$, $-1/2$ and $3/2$, and $e/3$ at the odd-denominator and hole-conjugate FQH states $\nu = -4/3$, $-1/3$ and $2/3$ respectively. We refer to this as the minimal excitation regime and mark it with open circles in Fig.~\ref{fig:figure_4}a,b. As the coupling is increased, we observe a transition to a double-gate-period regime, marked by filled circles: the ADG period doubles with respect to the minimal excitation value at $\nu = -5/2$, $-1/2$, $2/3$ and $3/2$, whereas no doubling is found at the odd-denominator FQH states $\nu = -4/3$ and $-1/3$ (see Supp. Material). This is similar to the observations reported in previous AD and interferometry experiments~\cite{Kou2012Jun, DiLuca2026Aug, Kim2026Jan, Henzinger2026Mar, Kim2026Mar}.

To understand the origin of the crossover between these two regimes, we consider how the side gates modify the system. As shown in Fig.~\ref{fig:figure_3}d,e for $\nu=2/3$, and similarly for $\nu=-1/2$ in Fig.~\ref{fig:figure_3}g,i, the doubling occurs over the same antidot-gate voltage range when only the side-gate voltage is varied. As illustrated in Fig.~\ref{fig:figure_3}a--c, increasing the side-gate voltage brings the extended edge states into closer proximity to the AD-bound states, thereby increasing their coupling. At the same time, the side gates modify the local confinement potential experienced by the extended edge states, which may alter its microscopic structure, including possible edge reconstruction \cite{Wen1994Mar, Wan2002Jan, Wang2013Dec, Sabo2017May}. We therefore consider two possible mechanisms for the observed crossover: quasiparticle bunching and a transition toward an Aharonov--Bohm-dominated. Both possibilities are schematically shown in Fig.~\ref{fig:figure_4}c.

One possible interpretation of the doubled ADG periodicity is quasiparticle bunching. Within a CD picture, this corresponds to the AD charge changing in increments of $2e^*$ rather than minimal excitation $e^*$. Effective charges larger than the minimal quasiparticle charge have already been reported at $\nu = 2/3$ and $\nu = 5/2$~\cite{Bid2009Dec, Dolev2010Apr}, where in the latter an $e/2$ agglomerate has been proposed as a competing low-energy
excitation~\cite{Carrega2011Sep}. Importantly in this scenario, the dominance of the $e/2$ process relies on interaction-induced renormalization of the edge dynamics, which may be relevant here since the side gates modify the AD--edge coupling and local electrostatic environment, and can therefore change the relative importance of competing quasiparticle processes. More recently, interferometers experiments has reported the observation of coherent pairing of $e/3$ at the hole-conjugate FQH state $\nu=2/3$~\cite{Ghosh2025Jun,Kim2026Jan, Henzinger2026Mar}. 

The appearance of the double-gate-period regime in the even-denominator and hole-conjugate FQH states studied here, but not at the conventional odd-denominator FQH states $\nu=-1/3$ and $-4/3$, suggests that the crossover depends on the nature and microscopic edge-structure of the underlying FQH state, and may be related to the presence of neutral modes in even-denominator and hole-conjugate states~\cite{Bid2010Jul, Gurman2012Dec, Dolev2011Jul}. We emphasize, however, that the doubled gate period alone does not establish microscopic quasiparticle bunching.

In fact, in this picture, adding two minimal quasiparticles per resonance would reduce the number of charge-addition events per flux quantum by a factor of two and therefore double the magnetic-field period. This is consistent with the behavior observed at $\nu=-5/2$ and $\nu=2/3$. However, bunching within this simple CD picture does not account for all of the data: at $\nu=3/2$ the magnetic-field periodicity corresponds to $p \approx 1$, while at $\nu=-1/2$ we obtain $p\approx1/3$, corresponding to an anomalously large magnetic-field period.

A quantitative interpretation of these deviations is presently difficult. Existing antidot theories largely consider idealized edge structures or weakly coupled sequential-tunneling and Coulomb-blockade limits~\cite{Merlo2007May,Sim2008Feb,Ilan2011Mar,LevySchreier2016Aug}, while separate theoretical works show that the edge structure of both hole-conjugate and even-denominator FQH states can be strongly modified by interactions and edge reconstruction~\cite{Wang2013Dec,Zhang2014Oct}. To our knowledge, there is currently no quantitative framework describing the gate- and magnetic-field periodicities of an AD in the presence of both strong AD-to-extended-edge coupling and a reconstructed, multimode fractional edge. The deviations from the simple $2e^*$ and $p/2$ picture therefore suggest that additional interaction-induced effects may become important as the coupling is increased.

We cannot, however, exclude a second interpretation in which the system crosses over away from the CD regime toward an AB-dominated or mixed transport regime. Theoretical work on quantum Hall ADs predicts that, in the AB limit, the flux period is one flux quantum ($\fieldratio=1$), independently of filling factor~\cite{LevySchreier2016Aug}. In this regime, the gate-voltage period no longer directly reflects the localized quasiparticle charge. For an ideal side gate that primarily changes the enclosed area without appreciably modifying the local density profile, and assuming that the ADG-to-area lever arm is approximately independent of filling factor, the normalized gate period in the AB limit is set by the filling factor, $\gateratio=\nu$~\cite{LevySchreier2016Aug}. Although the electrostatics of our gate-defined BLG antidot differ from this idealized model, tuning the side gates may drive the system away from the purely CD limit toward an intermediate or AB-dominated regime.

In our case the magnetic field periodicities at $\nu = -2/3$ and $\nu = 3/2$ fall close to $\fieldratio = 1$, as expected in the AB regime. For the other states the magnetic field periods do not approach $\phi_0$ as expected in the extreme AB limit. Additionally, the measured gate period ratios do not match the corresponding filling factors. They are instead consistent with recent AB interferometry experiments in BLG, where the dominant interfering charge is set by the filling of the partially occupied Landau level, $\nu_{\mathrm{LL}} = \nu - \lfloor \nu \rfloor$, rather than by the minimal quasiparticle charge~\cite{Kim2026Jan, Henzinger2026Mar}. Phenomenologically, the measured gate-period ratios show a striking correspondence with recent AB interferometry experiments in BLG: $\gateratio\approx1/2$ at $\nu=-5/2$, $-1/2$, and $3/2$, and $\gateratio\approx2/3$ at $\nu=2/3$, following the filling $\nu_{\mathrm{LL}}$ of the partially occupied Landau level~\cite{Kim2026Jan,Henzinger2026Mar,Kim2026Mar}. The observed crossover therefore cannot be described as a complete transition to this limit. Instead, it may indicate an intermediate or mixed regime in which interference, Coulomb interactions, and the microscopic edge structure all contribute to the observed periodicities.

The behavior at $\nu=-4/3$ and $-1/3$, however, shows that increasing the side-gate coupling does not generically drive the AD into the doubling regime. Even at the largest side-gate voltage accessible in our device, no period doubling is observed, and the magnetic-field periodicity remains close to $\phi_0/4$ for $\nu = -4/3$, rather than approaching $\phi_0$ (Supp. Material). This further supports that an AB crossover alone does not naturally explain why the doubled periodicity appears selectively in the even-denominator and hole-conjugate FQH states studied here. This state dependence suggests that the crossover may also involve the microscopic edge structure and the competition between different quasiparticle processes discussed above.

Finally, $\nu=-1/2$ remains anomalous within both pictures. Its magnetic-field period is substantially larger than at any other filling factor and cannot reasonably be attributed solely to a change in the effective antidot area, which would require a reduction of the enclosed area by close to a a factor of ten upon tuning only the side gates. One possibility is that this state lies in an intermediate transport regime, where Coulomb interactions and interference both contribute. In addition, $\nu=-1/2$ lies close to the highly resistive charge-neutrality point and is known to exhibit a complex edge structure, with several fractional plateaus emerging upon edge partitioning~\cite{Alkalai2026Feb}. We therefore cannot exclude additional modifications of the edge structure or the formation of lower density fractional regions between the side gates and the AD, which could further alter the antidot electrostatics and charging energy~\cite{Hajigeorgiou2026Jun}.

The temperature dependence (see Supp. Material) provides an additional indication that the doubled periodicity at $\nu=-1/2$ differs from that of the other even-denominator states. At $\nu=-5/2$ and $3/2$, the $e/4$ oscillations and their doubled-period counterparts are suppressed on a similar temperature scale, disappearing at approximately \SI{60}{\milli\kelvin}. This common temperature scale is consistent with the double-gate-voltage regime originating from the same underlying $e/4$ quasiparticles. At $\nu=-1/2$, in contrast, the doubled-period oscillations remain visible up to \SI{120}{\milli\kelvin}, suggesting that a different mechanism is involved.

The DC-bias dependence reported in the Supp. Material also does not provide conclusive information, because, even though the shape resembles Coulomb diamonds in both regimes, it has been shown that the coupling between the AD-bound states and the extended edge states can produce a different shape (diamond-like or checkerboard pattern-like) while the system remains in the same regime~\cite{Moreau2022Mar}. 

\section*{Conclusion}

In conclusion, we have measured reported oscillation periods consistent with charge-$e/4$ addition at $\nu = -5/2$, $-1/2$ and $3/2$, and with charge-$e/3$ addition at the hole-conjugate state $\nu = 2/3$. These measurements establish that the expected elementary fractional charge can be accessed and localized in a gate-defined AD in bilayer graphene when operated in the appropriate Coulomb-dominated regime.

By tuning the side gates we further observe a crossover from the minimal-excitation period to a doubled gate period. This crossover may reflect a change in the dominant charge-addition process, such as the bunching of two minimal excitations, or a transition towards an Aharonov-Bohm-dominated regime. Its microscopic origin remains an open question, since the side gates simultaneously modify the tunnel and electrostatic coupling, the confinement profile and, through edge reconstruction, the microscopic edge structure itself. The selective appearance of the doubled periodicity nonetheless provides an additional probe of how the quasiparticle dynamics evolve as the AD is coupled more strongly to the surrounding edge states.

More broadly, the ability to localize and resolve individual charge-$e/4$ excitations is a prerequisite for the study of candidate non-Abelian states in bilayer graphene. In smaller antidots, where the single-level splitting exceeds the temperature scale, this platform could allow deterministic control of individual $e/4$ quasiparticles~\cite{Simon2000Jun} and open a route towards experiments probing their exchange statistics and, ultimately, controlled braiding~\cite{DasSarma2005Apr, Nayak2008Sep}.

\textbf{\begin{center}Acknowledgments\end{center}}
The authors acknowledge insightful discussions with Bert Halperin and Taige Wang that have improved the interpretation of the results. Additionally, the authors acknowledge Steven H. Simon, Philip Kim, and Tevz Lotric for discussions and for comments about the manuscript. The authors acknowledge Paul Dyson for the additional financial support for the liquid helium. M.D.L. acknowledges Punam Barman for the exfoliation of the BLG flake used in this work, and acknowledges Blender for its open-source software. 

\textbf{\begin{center}Funding Statement\end{center}}
M.D.L., E.H. and N.M. acknowledge funding from SNSF. M.B. acknowledges the support of the SNSF Eccellenza grant No. PCEGP2\_194528, and support from the QuantERA II Program that has received funding from the European Union’s Horizon 2020 research and innovation program under Grant Agreement No 101017733. K.W. and T.T. acknowledge support from the JSPS KAKENHI (Grant Numbers 20H00354 and 23H02052) and World Premier International Research Center Initiative (WPI), MEXT, Japan. 

\textbf{\begin{center}Author contributions\end{center}}
M.B. supervised the project. M.D.L. fabricated the devices. M.D.L. and N.M. developed the air-bridge recipe. E.H. improved and installed the filters on the cold finger, which enabled a lower electron temperature. E.H. and M.D.L performed the measurements. M.D.L, E.H., and A.W. analyzed the data. M.D.L and E.H. wrote the manuscript with input from all authors. T.T. and K.W. provided the hBN crystal. 

\textbf{\begin{center}Competing Interests\end{center}}
The authors declare no competing interests.

\section*{METHODS}

\subsection{Sample fabrication}
The device was fabricated using a standard van der Waals dry-transfer technique. First, graphene and hBN were mechanically exfoliated from bulk crystals on a SiO2/Si substrate. The desired flakes are identified under an optical microscope and an atomic force microscope to check for any flake impurity. The stack was assembled using homemade poly(bisphenol A carbonate)/polydimethylsiloxane (PC/PDMS) to pick up all the flakes at a temperature of approximately 90 $^\circ$C. The stack was then transferred onto a doped silicon substrate with a 285 nm thick layer of thermally grown SiO$_2$ by melting the PC at a temperature of approximately 180 $^\circ$C. The measured stack has a 46\,nm top hBN layer and a 20\,nm bottom hBN layer.

Device patterning was achieved through multiple steps of electron beam lithography, followed by reactive ion etching and metal deposition. First, the TG contacts were created by depositing a 5/30\,nm Cr/Au layer. The device geometry was then defined through reactive ion etching, initially using O$_2$, followed by SF$_6$, and then another O$_2$ step. The electrode pattern was subsequently created by etching with SF$_6$ and O$_2$, followed by angled deposition of a 5/15/70\,nm Cr/Pd/Au layer with rotation. The TG was refined with 15-second O$_2$ etch steps. During each step, the two-probe resistance between each gate and contact was monitored to ensure that all gates were fully separated while minimizing etching of the hBN and achieving narrow line widths~\cite{Ronen2021May}. The air-bridge was fabricated as a last step using a 3-layer resist stack composed of PMMA 950K A4, MMA EL9, PMMA 495K A4. The dose is precisely calibrated such that at the foot regions of the air-bridge all three resist layers are fully exposed, while in the flying region only the top two layers are exposed, leaving the bottom PMMA 950K A4 layer unexposed and intact after development. Subsequently, 25/250\,nm of Cr/Au are evaporated, and the liftoff is performed in acetone, which also dissolves the resist beneath the flying region, leaving the metal air-bridge suspended over the device.\\

\subsection{Measurements}
The device was measured in a Leiden wet dilution refrigerator with a base temperature of approximately $\qty{10}{\milli\kelvin}$. Electronic filters were installed on all transport lines to improve electron thermalization. Resistance measurements were performed using Zurich Instruments MFLI lock-in amplifiers at a frequency of $\qty{17.777}{\hertz}$. An AC voltage excitation of $\qty{0.2-1}{\volt}$ was applied through a $\qty{1}{\giga\ohm}$ bias resistor, corresponding to a current excitation of $\qty{0.2-1}{\nano\ampere}$. For each filling factor, we selected a current sufficiently low to avoid merging distinct features or activating additional transport paths and unwanted excitations, while remaining high enough to provide an adequate signal-to-noise ratio. The voltage signals were amplified at room temperature by a factor of 100 using LI-75A low-noise voltage pre-amplifiers before being sent to the lock-in amplifiers. For Coulomb-diamond measurements, the presence of the cold ground prevented a direct measurement of the current. We therefore applied a $\qty{10}{\micro\volt}$ AC excitation together with a DC-voltage bias supplied by a Quantum Machines QDAC-II ultra-low-noise 24-channel DAC, through a homemade voltage adder. Since the current was not measured in this configuration, the resistance could not be extracted; for these DC-bias measurements, we instead report the voltage measured across the AD, $V_{\mathrm{t}}$. Gate voltages were applied using the QDAC-II. To improve the quality of the ohmic contacts, a silicon-gate voltage of $V_{\mathrm{Si}}=+\qty{40}{\volt}$ was applied for electron-like states, while $V_{\mathrm{Si}}=-\qty{40}{\volt}$ was applied for hole-like states. This dopes the graphene near the contacts into highly electron- or hole-doped regions, respectively.

\subsection{Period and error estimation of the oscillation}
\label{sec:error}

For each filling factor, the 2D-FFT provides information about the gate-voltage and magnetic-field periods. To suppress low-frequency features arising from the finite measurement window and slowly varying backgrounds, a third-order polynomial background is subtracted from the raw data before performing the 2D-FFT. In addition, to compensate for the magnetic field instability of the power supply, we apply a two-point moving average along the magnetic field axis. Both the raw and background-subtracted data are shown in the Supp. Material for all data included in the paper.

To extract the characteristic periods, we fit each horizontal line of the 2D-FFT with a Gaussian and select the line with the largest fitted amplitude, which identifies the dominant peak in the gate-voltage direction. A vertical line cut through this peak is then fitted with a Gaussian to determine the magnetic-field frequency. The fitted peak positions are converted to the corresponding periods, $\Delta V$ and $\Delta B$. The extracted periodicities are also verified by visual inspection of the conductance maps.

As an independent cross-check, we identify peaks directly in the conductance maps and compile their peak-to-peak spacings into histograms. Gaussian fits to these distributions provide independent estimates of $\Delta V$ and $\Delta B$. We generally use the mean values extracted from the 2D-FFT analysis. However, when the uncertainty obtained from the 2D-FFT fit is not considered reliable, for example due to noise, instability, or proximity to zero frequency, we instead use the uncertainty obtained from the corresponding histogram analysis in order to avoid underestimating the error. Such cases are explicitly indicated in the Supp. Material.

The error obtained in this way accounts for both the measurement uncertainty and the statistical spread of the oscillation periods. For Fig.~\ref{fig:figure_4}a we add a further contribution, to account for the change in the effective AD area between the reference state at $\nu = -1$ and the state of interest. We estimate this change by assuming that each additional edge state displaces the AD-bound states by one magnetic length $l_B$. At $B = \SI{14}{\tesla}$, $l_B = \SI{6.9}{\nano\metre}$, so for an AD of diameter \SI{360}{\nano\metre} this corresponds to an area variation of approximately \SI{8}{\percent}. We therefore include an additional \SI{8}{\percent} uncertainty in the error propagation when taking ratios of gate voltage periods to determine $\fieldratio$.

Most of the measurements presented in the main text were acquired with a magnetic-field step of $dB=1$\,mT, the minimum step reliably provided by our magnet power supply, with typically $N_\mathrm{B}\approx101$ points per scan. The frequency resolution of the FFT is determined by the total magnetic-field range and is approximately $\delta f_\mathrm{B}\simeq 1/(N_\mathrm{B}dB)$. In addition, the discrete field step limits the resolution of short magnetic-field periods. For states with small $\Delta B$, only a few measurement points are acquired per oscillation period, leading to a larger relative uncertainty in the extracted periodicity. This sampling limitation accounts for the larger error bars observed for the smallest magnetic-field periods.

\textbf{\begin{center}Data availability\end{center}}
The data supporting the findings of this study are available from the corresponding author upon reasonable request.

\bibliography{biblio}

\onecolumngrid
\newpage

\clearpage

\section*{Supplementary Material}

\subsection{Device Characterization}

\begin{figure*}[tph!]
 \renewcommand{\thefigure}{S1}
 \includegraphics[width = 0.95\textwidth]{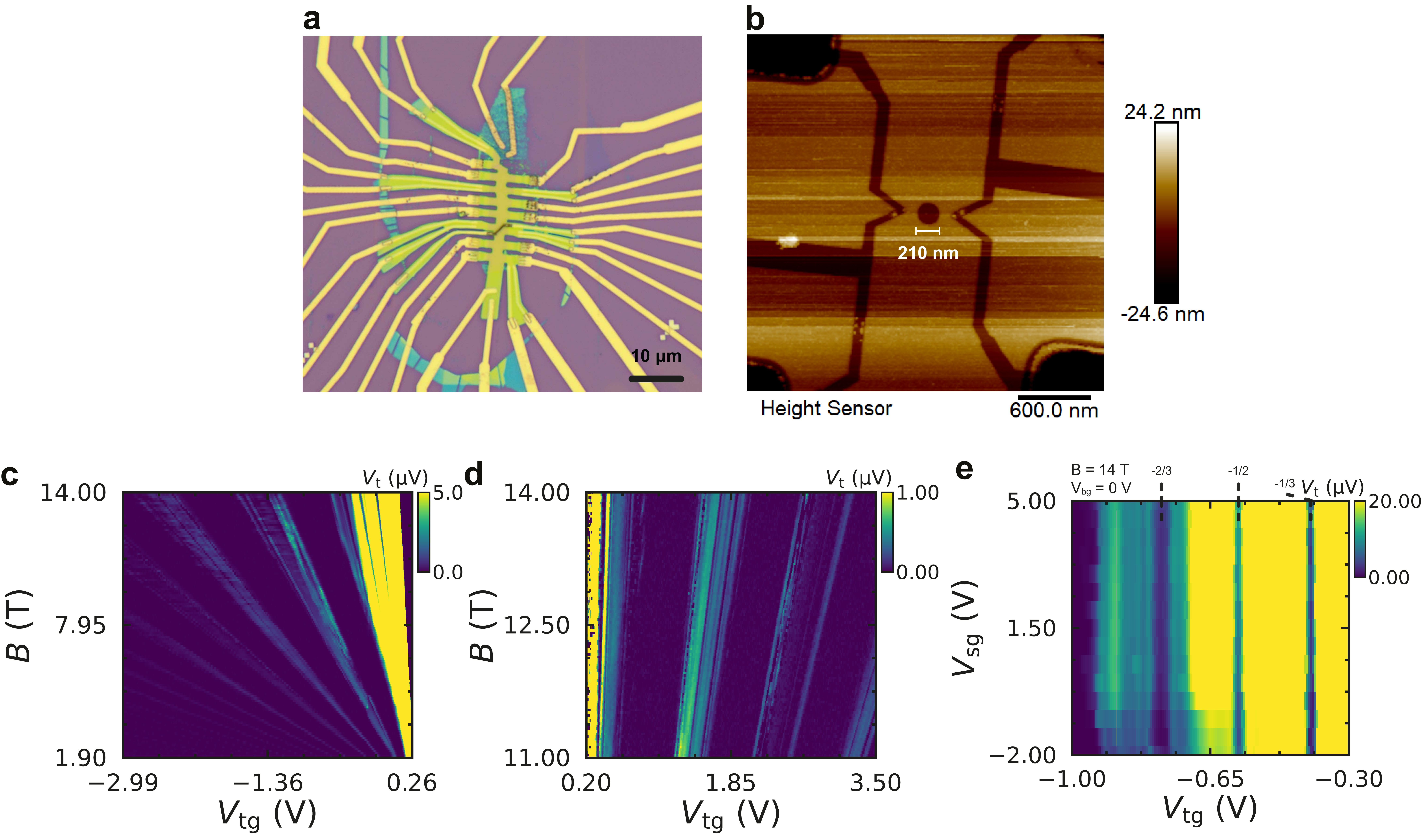}
 \begin{center}
\caption{\textbf{Device characterization}. \textbf{a}, Optical microscope image of the device. Scale bar is \SI{10}{\micro\meter}. \textbf{b}, Atomic force microscopy image of the device taken after the etching of the top graphite gate and before the air-bridge step. The hole defining the AD, the trenches defining the side gates and the region where the air-bridge will be deposited are visible. After etching, the AD diameter is approximately \SI{210}{\nm}. \textbf{c--d}, Transmitted voltage $V_\mathrm{t}$ across the AD as a function of the top gate voltage, $V_\mathrm{tg}$, and the magnetic field, $B$, on the hole (\textbf{c}) and electron (\textbf{d}) side respectively, showing the development of the Landau levels. \textbf{e}, Example of characterization of the side gates: transmitted voltage as a function of $V_\mathrm{tg}$ and the side gate voltage, $V_\mathrm{sg}$. On the hole side, as the side gate voltage is increased, the voltage drop across the AD increases, indicating that the extended-edge states are more strongly coupled to the AD and that backscattering is enhanced. The side gates are swept together. In \textbf{c--e} we report $V_\mathrm{t}$ instead of the resistance, since the measurements were taken in a voltage source configuration.}
 \label{fig:Characterization}
 \end{center}
\end{figure*}

An important part of the characterization is the tuning of the side gates. Figure~\ref{fig:Characterization}e shows an example of this procedure: the transmitted voltage is measured as a function of the top gate voltage $V_\mathrm{tg}$ and the side gate voltage $V_\mathrm{sg}$, with the two side gates swept together. On the hole side, as $V_\mathrm{sg}$ is increased, the voltage drop across the AD increases, indicating that the extended edge states are more strongly coupled to the AD and that backscattering is enhanced. We note that the transmission of the edge states is only weakly affected by the side gates: even over a large voltage range these states are not completely pinched off. The accessible range of $V_\mathrm{sg}$ is limited by the risk of shorting the side gates to the top gate, which sets an upper bound on the voltage we can safely apply, in our case of approximately 6\,V.

\clearpage

\subsection{Capacitance calculations for ADG periodicity}

\begin{figure*}[tph!]
 \renewcommand{\thefigure}{S2}
 \includegraphics[width = 0.95\textwidth]{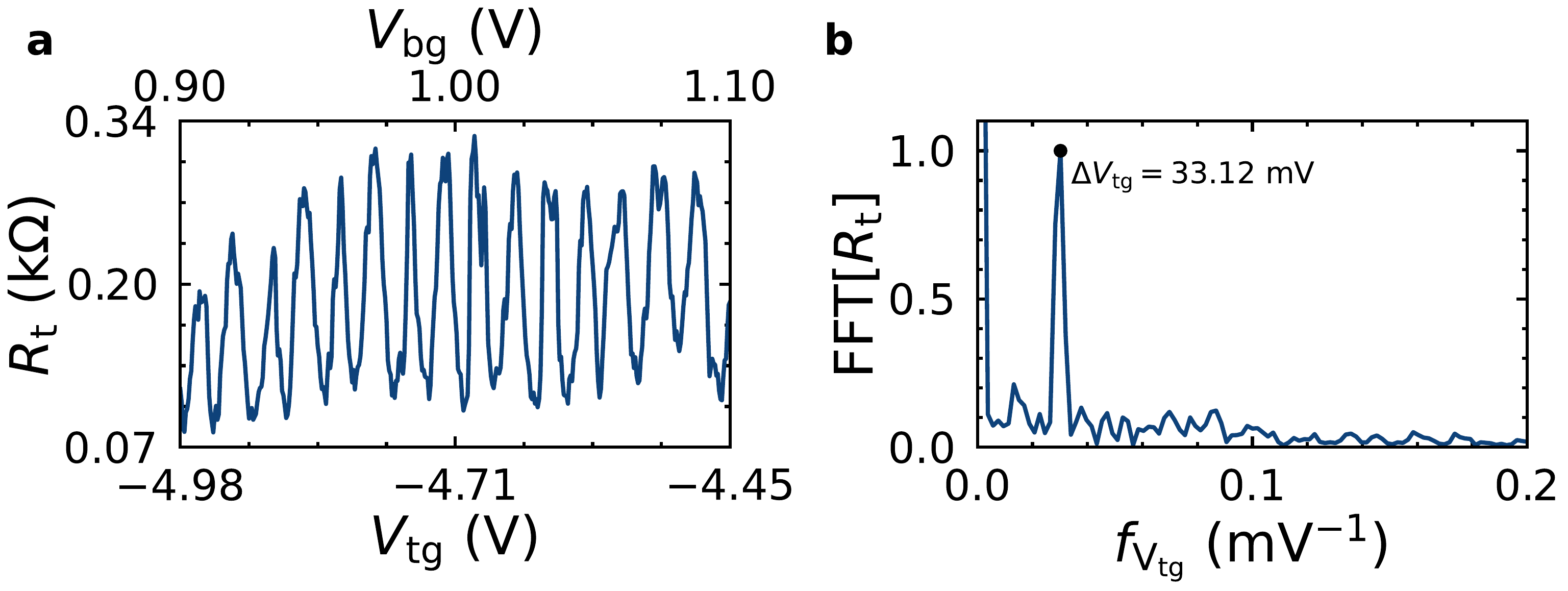}
 \begin{center}
\caption{\textbf{Area oscillations}. \textbf{a}, Transmitted resistance $R_\mathrm{t}$ oscillations at $\nu = -2$ as a function of the top gate voltage $V_\mathrm{tg}$ and the back gate voltage $V_\mathrm{bg}$ when they are sweeped simultaneously. The oscillations are obtained by keeping the bulk density constant and varying only the displacement field, and correspond to area oscillations in which the displacement field modifies the antidot potential, similar to the effect of the ADG. \textbf{b}, One-dimensional FFT of the trace in \textbf{a}, from which the periodicity in $V_\mathrm{tg}$ is extracted.}
 \label{fig:Area}
 \end{center}
\end{figure*}

The ADG primarily controls the shape of the AD potential. Its periodicity can therefore be understood through a simple capacitive model, in which the ADG and the top graphite gate act as two parallel-plate capacitors in series with the graphene,

\begin{align}
 C_\mathrm{adg} &= \left(C_\mathrm{tg}^{-1} + C_\mathrm{air}^{-1}\right)^{-1}\nonumber\\
 &= C_\mathrm{tg}\left(1 + \epsilon_\mathrm{hBN}\frac{d_\mathrm{bridge}}{d_\mathrm{tg}}\right)^{-1},
 \nonumber
\end{align}

\noindent where $\epsilon_\mathrm{hBN}$ is the dielectric constant of hBN, $d_\mathrm{tg}$ the thickness of the top hBN, and $d_\mathrm{bridge}$ the height of the air-bridge above the top hBN. The air-bridge height cannot be determined precisely: from fabrication, we expect the suspended part to sit at approximately \qty{160}{\nm} (which is the thickness of the bottom PMMA 950K A4 layer), though this value can vary significantly. Using a top hBN thickness of \qty{50}{\nm} and $\epsilon_\mathrm{hBN} \approx 3$, the model predicts $C_\mathrm{adg} \approx C_\mathrm{tg}/10$. The gate voltage period scales as the inverse of the capacitance, so the geometric capacitance ratio predicts $\Delta V_\mathrm{adg} \approx 10\,\Delta V_\mathrm{tg}$. This value is consistent with area oscillations, induced by keeping the bulk density constant and sweeping the TG and BG together. We confirm this directly using the area-oscillation shwon in Fig.~\ref{fig:Area}a at $\nu = -2$, which at fixed density gives a TG period of approximately \qty{33}{\mV}, Fig.~\ref{fig:Area}b. Combined with the geometric factor of 10, this gives a predicted ADG period of approximately \qty{330}{\mV}, closer to the measured value of approximately $\qty{235}{\mV}$. This agreement confirms that the ADG oscillations are area oscillations rather than density oscillations.

\clearpage

\subsection{Integer $\Delta V_1$, $\Delta B_1$ period determination}
\label{sec:period_determination}

\begin{figure*}[tph!]
 \renewcommand{\thefigure}{S3}
 \includegraphics[width = 0.7\textwidth]{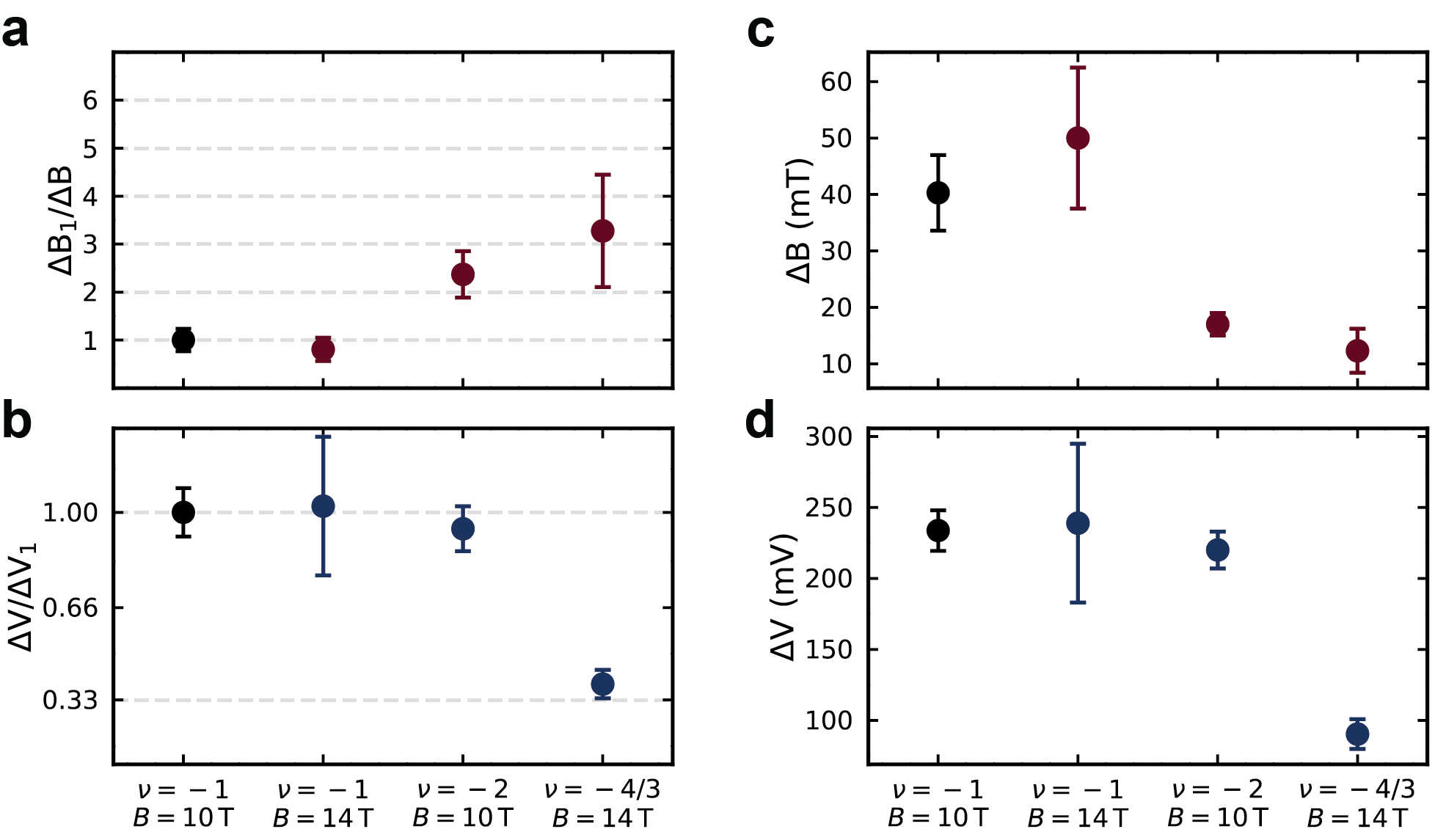}
 \begin{center}
 \caption{\textbf{Determination of the reference periods $\Delta V_1$ and $\Delta B_1$.} The black data point in each panel, corresponding to electron oscillations at $\nu=-1$ and $B=10$\,T, is used as the reference throughout. \textbf{a}, Magnetic-field period ratio, $\fieldratio$, normalized to the reference point. \textbf{b}, Antidot gate-voltage period ratio, $\gateratio$, normalized to the same reference point. \textbf{c,d}, Extracted magnetic-field and gate-voltage periods, respectively, for the four data points.}
 \label{fig:sup_integer_summary}
 \end{center}
\end{figure*}

Throughout the manuscript, we define $\Delta V_1$ as the antidot-gate voltage period corresponding to the addition of a single electron, which is the expected voltage periodicity at integer filling factors. Similarly, $\Delta B_1$ denotes the magnetic-field period associated with the addition of one flux quantum, $\phi_0$. To establish these reference scales, we therefore select an integer filling factor from which $\Delta V_1$ and $\Delta B_1$ are determined and subsequently use them to normalize the oscillation periods measured at other filling factors.

\begin{figure*}[tph!]
 \renewcommand{\thefigure}{S4}
 \includegraphics[width = 0.95\textwidth]{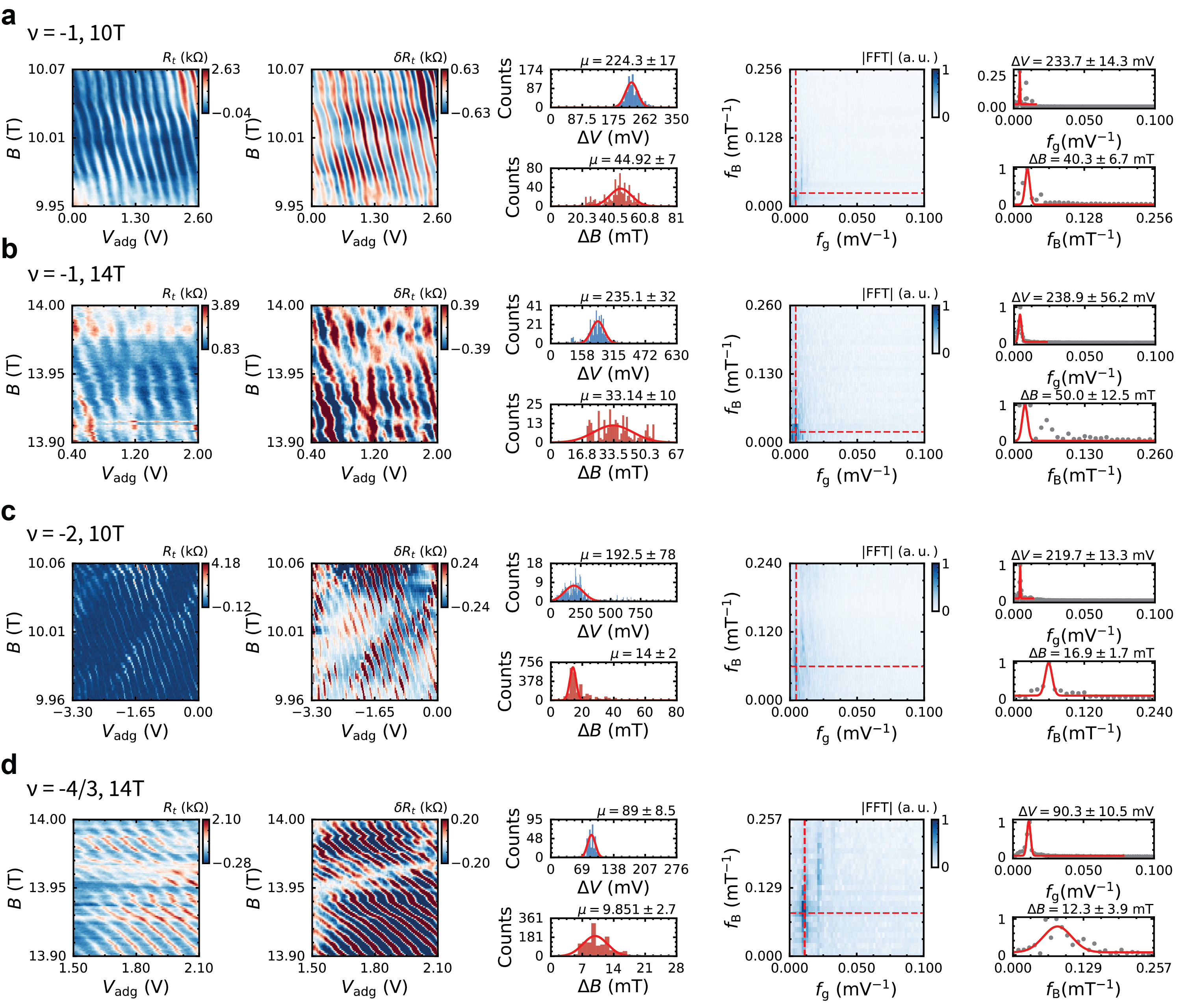}
 \begin{center}
\caption{\textbf{Oscillations for the states used to verify the calibration analysis.} For each row, the panels show, from left to right, the raw data; the data after background subtraction; histograms of the extracted peak-to-peak spacings in gate voltage, $\Delta V$, and magnetic field, $\Delta B$; the 2D-FFT; and line cuts through the 2D-FFT along the red dashed lines. The background is obtained by fitting each row of the raw data with a third-order polynomial and subtracting the fit, thereby removing the slowly varying background and suppressing the corresponding low-frequency components in the 2D-FFT. The distributions of $\Delta V$ and $\Delta B$ are fitted with Gaussian functions, with the Gaussian means providing estimates of the characteristic oscillation periods. The 2D-FFT is performed on the background-subtracted data, and the corresponding line cuts are also fitted with Gaussian functions. The Gaussian peak positions provide an independent determination of the gate-voltage and magnetic-field periodicities, while their widths characterize the spread of the corresponding peaks. Analysis of data in \textbf{a}, $\nu = -1$ at $B = 10$\,T, \textbf{b}, $\nu = -1$ at $B = 14$\,T, \textbf{c}, $\nu = -2$ at $B = 10$\,T and \textbf{d}, $\nu = -4/3$ at $B = 14$\,T.}
 \label{fig:sup_integer}
 \end{center}
\end{figure*}

As shown in Fig.~\ref{fig:sup_integer}b, the $\nu=-1$ data at $14$\,T exhibit substantially stronger fluctuations than at lower magnetic field, see Fig.~\ref{fig:sup_integer}a. The corresponding extracted gate-voltage and magnetic-field periods are shown in Fig.~\ref{fig:sup_integer_summary}c,d and therefore carry large uncertainties for $\nu=-1$ at $14$\,T. Since the integer reference is used to calibrate both the electron charge periodicity and the flux periodicity $\phi_0$, we require a reference point for which both periods can be extracted with relatively small error. We therefore use $\nu=-1$ at $10$\,T as the integer reference. This is the same method used in interferometry measurements where the area of the interferometer, and so the magnetic field periodicity is set by oscillations in the integer quantum Hall regime, generally measured at a different magnetic field than those in the FQH regime~\cite{Ghosh2025Jul, Samuelson2026Mar}. 

The consistency of this calibration is summarized in Fig.~\ref{fig:sup_integer_summary}. The $\nu=-1$ measurements at both $10$\,T and $14$\,T are consistent with electron charge periodicity and with $p=1$, corresponding to a magnetic-field periodicity of $\phi_0$. Using the more precise $\nu=-1$ data at $10$\, as the reference, with $\Delta V_1=233\pm14.3$\,mV and $\Delta B_1=40.3\pm6.7$\,mT, we further find that the $\nu=-2$ gate-voltage periodicity is consistent with electron charge, while its magnetic-field periodicity is consistent with $p=2$, corresponding to $\phi_0/2$, as expected for two edge states in the CD regime.

We further test whether the reference periods $\Delta V_1$ and $\Delta B_1$, determined at $10$\,T, can be reliably applied to measurements performed at $14$\,T. For this purpose, Fig.~\ref{fig:sup_integer_summary} shows the corresponding analysis for the fractional state at $\nu=-4/3$ at $B = 14$\,T. Previous AD measurements in the CD regime at this filling factor \cite{DiLuca2026Aug} found a magnetic-field periodicity corresponding to $p=4$, i.e., four quasiparticles added per flux quantum, together with a localized quasiparticle charge of $e/3$. Applying the $\nu=-1$ reference periods to our $\nu=-4/3$ data yields results consistent with both $p=4$ and charge $e/3$. This agreement provides an independent check that the reference periods determined at $10$\,T can be applied to the $14$\,T measurements. We therefore use $\Delta V_1$ and $\Delta B_1$ defined from the $\nu=-1$ oscillations throughout this work. The data underlying Fig.~\ref{fig:sup_integer_summary} are shown in Fig.~\ref{fig:sup_integer}.

\clearpage

\subsection{Oscillations using the TG for $\nu = -1$ and $-4/3$}

\begin{figure*}[tph!]
 \renewcommand{\thefigure}{S5}
 \includegraphics[width = 0.95\textwidth]{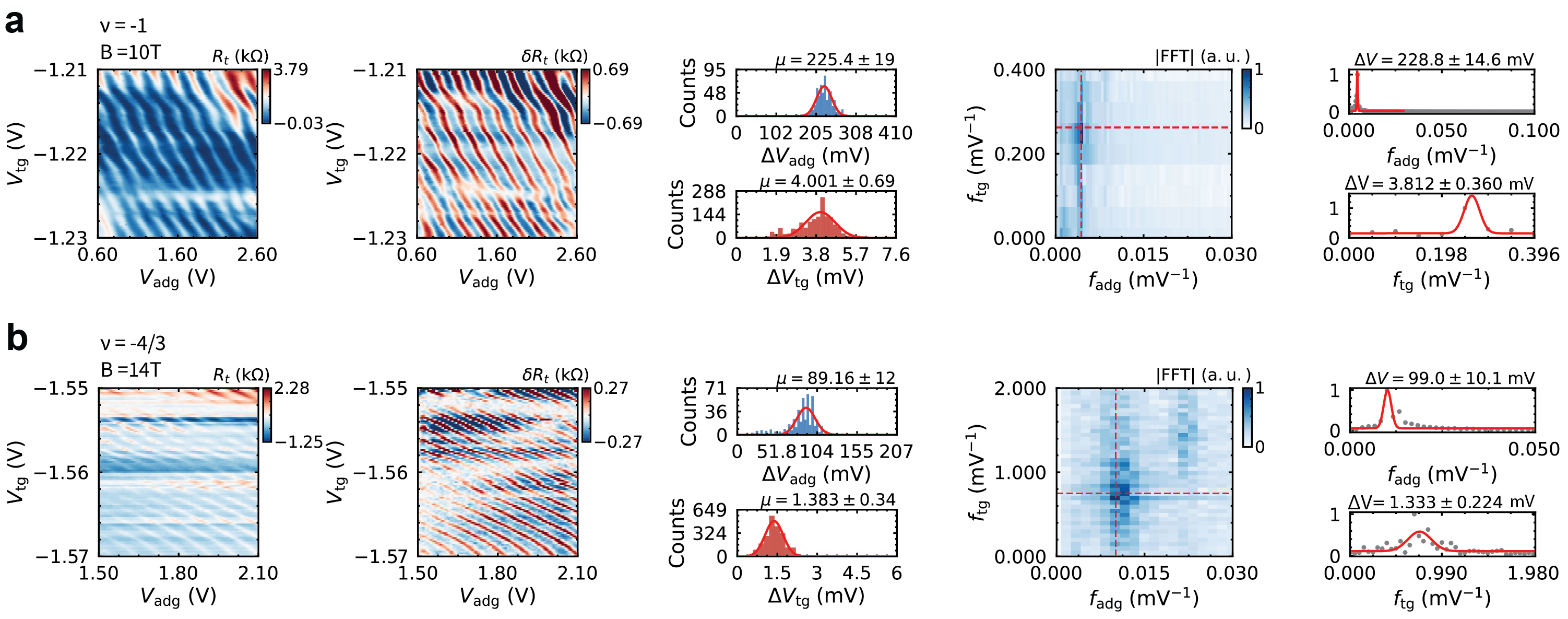}
 \begin{center}
\caption{\textbf{Antidot gate versus top gate maps.} For each row, the panels show, from left to right, the raw data; the data after background subtraction; histograms of the extracted peak-to-peak spacings in gate voltage, $\Delta V$, and magnetic field, $\Delta B$; the 2D-FFT; and line cuts through the 2D-FFT along the red dashed lines. The background is obtained by fitting each row of the raw data with a third-order polynomial and subtracting the fit, thereby removing the slowly varying background and suppressing the corresponding low-frequency components in the 2D-FFT. The distributions of $\Delta V$ and $\Delta B$ are fitted with Gaussian functions, with the Gaussian mean providing estimate of the characteristic oscillation period. The 2D-FFT is performed on the background-subtracted data, and the corresponding line cuts are also fitted with Gaussian functions. The Gaussian peak positions provide an independent determination of the gate-voltage and magnetic-field periodicities, while their widths characterize the spread of the corresponding peaks. \textbf{a}, ADG versus top gate for $\nu = -1$ at $B = 10$\,T. \textbf{b}, ADG versus top gate for $\nu = -4/3$ at $B = 14$\,T.}
 \label{fig:TGvsADG}
 \end{center}
\end{figure*}

To demonstrate that the ADG can be used equivalently to the top gate to extract the quasiparticle charge, for the data shown in Fig.~\ref{fig:figure_1} we sweep the ADG together with the TG and compare $\nu=-1$ at $B = 10$\,T with $\nu=-4/3$ at $B = 14$\,T. Both gates produce oscillations, though of a different nature. The TG produces density oscillations \cite{DiLuca2026Aug}, while the ADG oscillations occur from changes in the AD area.

If electrons tunnel at $\nu=-1$ and quasiparticles with charge $e/3$ tunnel at $\nu=-4/3$, the corresponding top-gate periodicities are expected to differ by a factor of three, such that $\Delta V_{\mathrm{tg}}(\nu=-1)/\Delta V_{\mathrm{tg}}(\nu=-4/3) = 3$. When using the periods extracted from the 2D-FFT, find $\Delta V_{\mathrm{tg}}(\nu=-1)/\Delta V_{\mathrm{tg}}(\nu=-4/3) = 2.9 \pm 0.6$, consistent with the expected ratio. This is the same principle used in our previous related work, where the quasiparticle charge was inferred using only the top-gate periodicity~\cite{DiLuca2026Aug}.

\clearpage

\subsection{Absence of doubling at $\nu = -4/3$}
\label{sec:absence_m43}

\begin{figure*}[tph!]
 \renewcommand{\thefigure}{S6}
 \includegraphics[width = 0.95\textwidth]{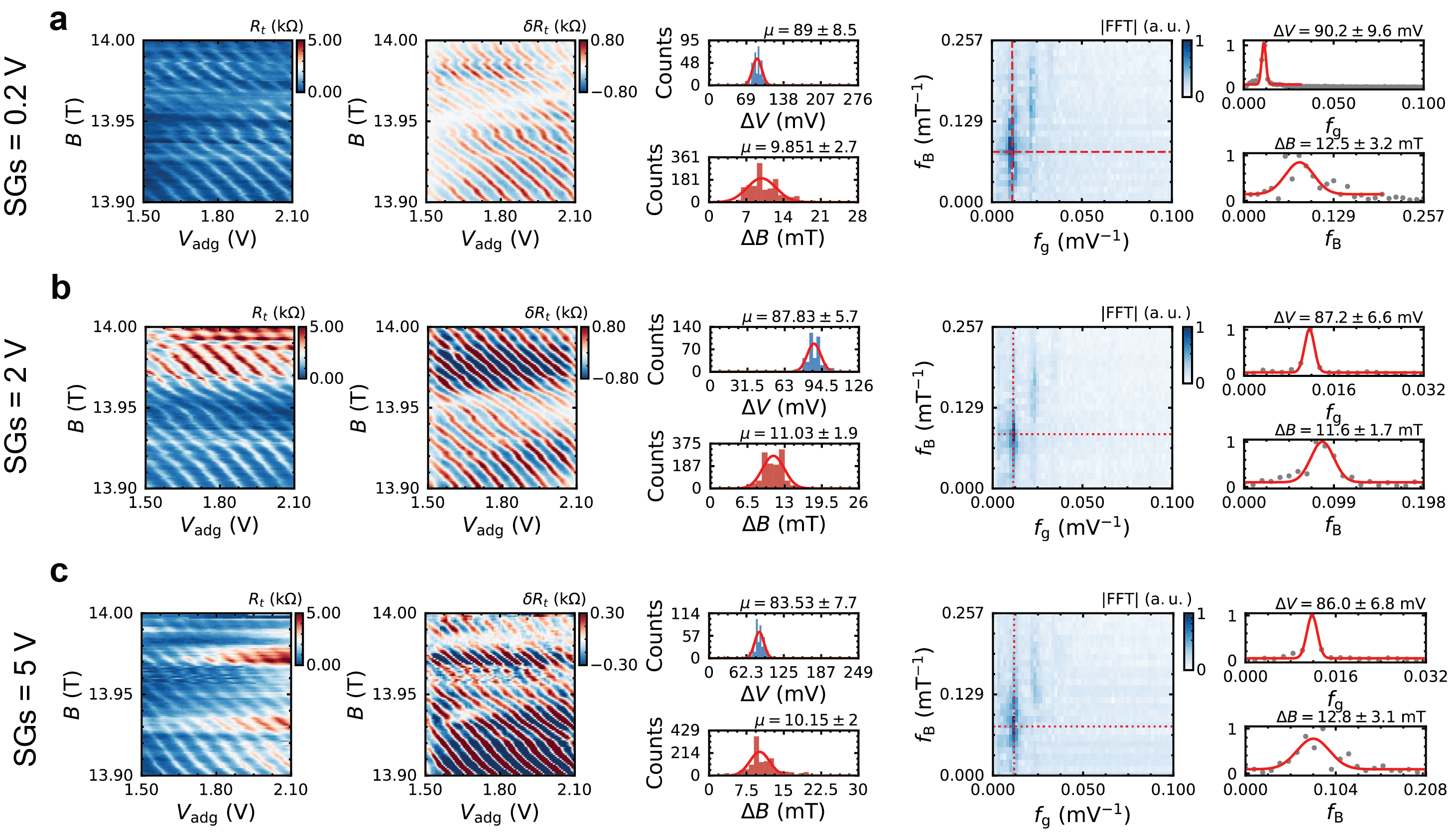}
 \begin{center}
\caption{\textbf{Oscillations at $\nu=-4/3$ as a function of the side gate voltage.} \textbf{a--c}, Measurements performed for side-gate voltages of $0.2$\,V, $2$\,V, and $5$\,V, respectively, while all other gate voltages are kept fixed. A more positive side-gate voltage increases the coupling between the extended edge state and the AD-bound states. In all three side-gate configurations, we extract a charge consistent with $e/3$ and a magnetic-field periodicity corresponding to $\phi_0/4$ ($p=4$). No transition to the Aharonov--Bohm regime is observed, or doubling of the gate voltage period, even at the maximum side-gate voltage of $5$\,V allowed by the safe limits of the device. For each side-gate configuration, the panels show, from left to right, the raw data; the data after background subtraction; histograms of the extracted peak-to-peak spacings in gate voltage, $\Delta V$, and magnetic field, $\Delta B$; the 2D-FFT; and line cuts through the 2D-FFT along the red dashed lines. The background is obtained by fitting each row of the raw data with a third-order polynomial and subtracting the fit, thereby removing the slowly varying background and suppressing the corresponding low-frequency components in the 2D-FFT. The distributions of $\Delta V$ and $\Delta B$ are fitted with Gaussian functions, with the Gaussian means providing estimates of the characteristic oscillation periods. The 2D-FFT is performed on the background-subtracted data, and the corresponding line cuts are also fitted with Gaussian functions. The Gaussian peak positions provide an independent determination of the gate-voltage and magnetic-field periodicities, while their widths characterize the spread of the corresponding peaks.}
 \label{fig:absence_m4_3}
 \end{center}
\end{figure*}

\clearpage

\subsection{Absence of doubling at $\nu = -1/3$}
\label{sec:absence_m13}

\begin{figure*}[tph!]
 \renewcommand{\thefigure}{S7}
 \includegraphics[width = 0.95\textwidth]{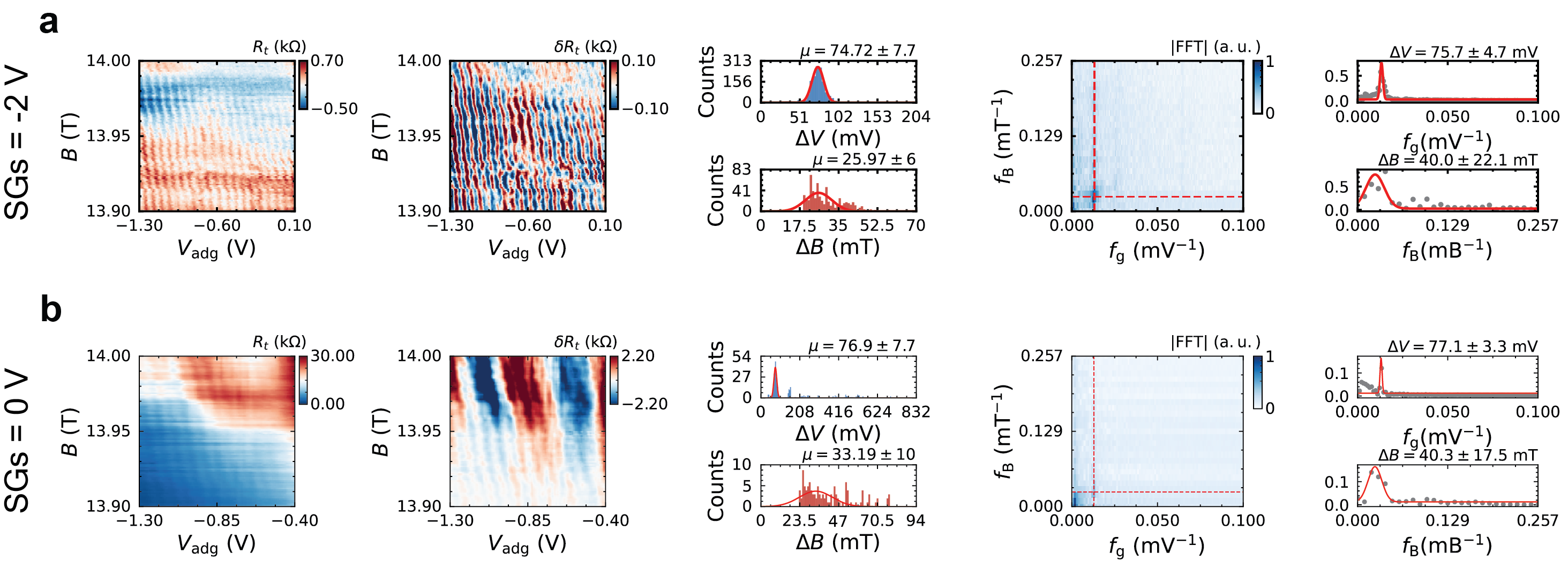}
 \begin{center}
\caption{\textbf{Oscillations at $\nu=-1/3$ as a function of the side gate (SGs) voltage.} \textbf{a--b}, Measurements performed for side-gate voltages of $-2$\,V, and $0$\,V, respectively, while all other gate voltages are kept fixed. A more positive side-gate voltage increases the coupling between the extended edge state and the antidot bound states. In both side-gate configurations, we extract a charge consistent with $e/3$ and a magnetic-field periodicity corresponding to $\phi_0$ ($p=1$). No doubling of the gate voltage periodicity is observed. For each side-gate configuration, the panels show, from left to right, the raw data; the data after background subtraction; histograms of the extracted peak-to-peak spacings in gate voltage, $\Delta V$, and magnetic field, $\Delta B$; 2D-FFT; and line cuts through the 2D-FFT along the red dashed lines. The background is obtained by fitting each row of the raw data with a third-order polynomial and subtracting the fit, thereby removing the slowly varying background and suppressing the corresponding low-frequency components in the 2D-FFT. The distributions of $\Delta V$ and $\Delta B$ are fitted with Gaussian functions, with the Gaussian means providing estimates of the characteristic oscillation periods. The 2D-FFT is performed on the background-subtracted data, and the corresponding line cuts are also fitted with Gaussian functions. The Gaussian peak positions provide an independent determination of the gate-voltage and magnetic-field periodicities, while their widths characterize the spread of the corresponding peaks.}
 \label{fig:absence_m1_3}
 \end{center}
\end{figure*}

\clearpage

\subsection{Additional data for the Coulomb-dominated oscillations}

\begin{figure*}[tph!]
 \renewcommand{\thefigure}{S8}
 \includegraphics[width = 0.8\textwidth]{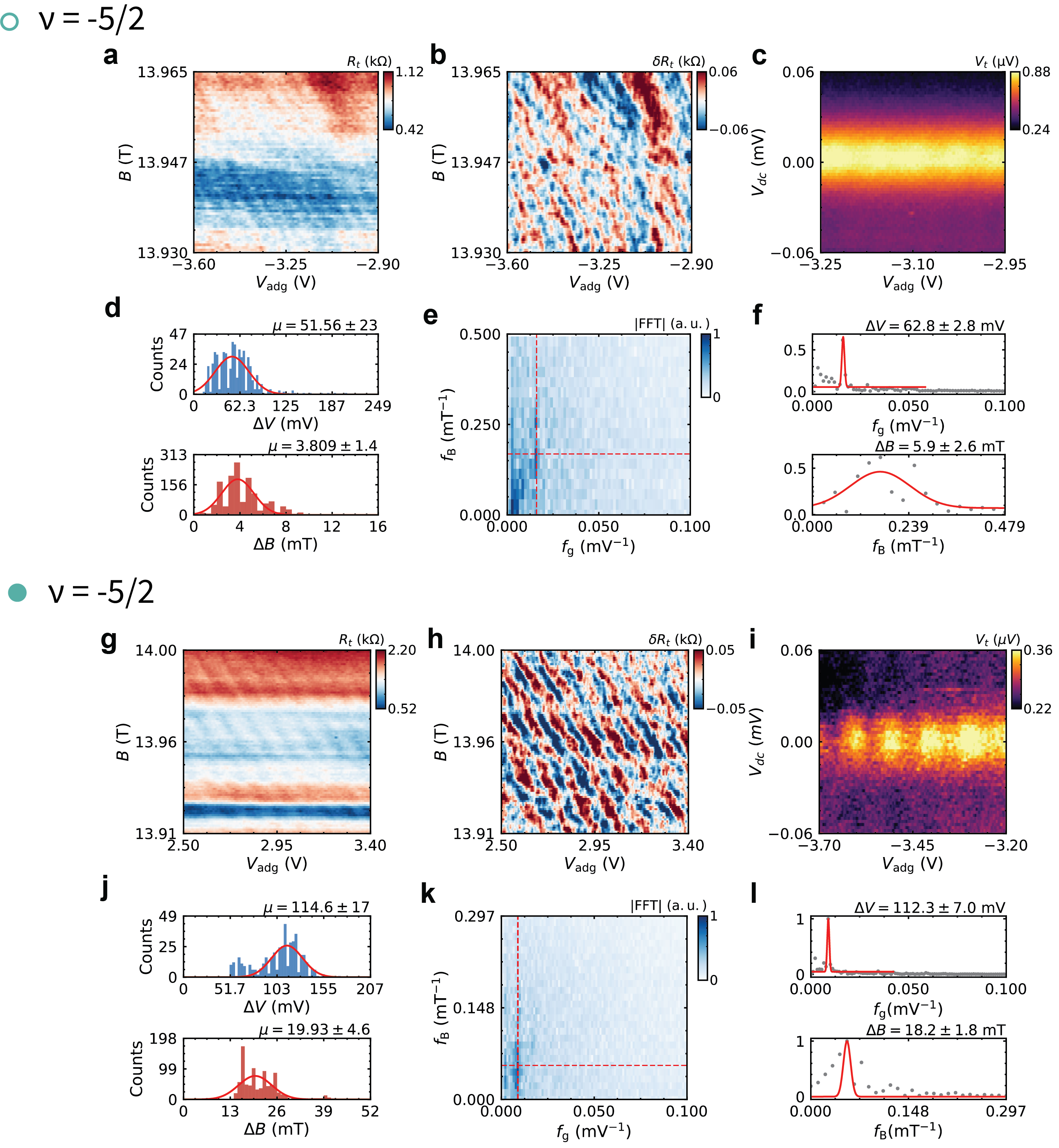}
 \begin{center}
 \caption{\textbf{Full dataset for $\nu = -5/2$.} Panels \textbf{a--f} correspond to the minimal excitation, while panels \textbf{g--l} to the double gate period regime. \textbf{a,g}, Raw $V_\mathrm{adg}$ versus $B$ map. \textbf{b,h}, The data after background subtraction. \textbf{c,i}, DC-bias dependence of oscillations, measuring $V_\mathrm{t}$ as explained in Methods. \textbf{d,j}, Histograms of the extracted peak-to-peak spacings in gate voltage, $\Delta V$, and magnetic field, $\Delta B$. \textbf{e,k}, 2D-FFT of the oscillations, and \textbf{f,l}, are line cuts through the 2D-FFT along the red dashed lines. For the doubling-regime data, inspection of the line cut and corresponding fit along the vertical dashed line in \textbf{l} shows that the Gaussian does not provide a satisfactory description of the peak, suggesting that the uncertainty obtained from the fit is underestimated. We therefore retain the Gaussian mean as the extracted value of $\Delta B$, but assign the uncertainty obtained from the $\Delta B$ histogram in order to avoid underestimating the error. At $\nu = -5/2$ the minimal excitation oscillations are considerably weaker than at the other fillings, and the corresponding 2D-FFT peak is faint. The fit of this peak returns an uncertainty on the gate voltage period of \SI{2.8}{\milli\volt}, half of what we obtain at states with comparable periods but stronger oscillations. We consider this value not representative of the actual confidence on the period, and we therefore adopt a more conservative uncertainty of \SI{5.6}{\milli\volt}, in line with the other states. This is the value reported in the main text.}
 \label{fig:sup_m52}
 \end{center}
\end{figure*}

\begin{figure*}[tph!]
 \renewcommand{\thefigure}{S9}
 \includegraphics[width = 0.8\textwidth]{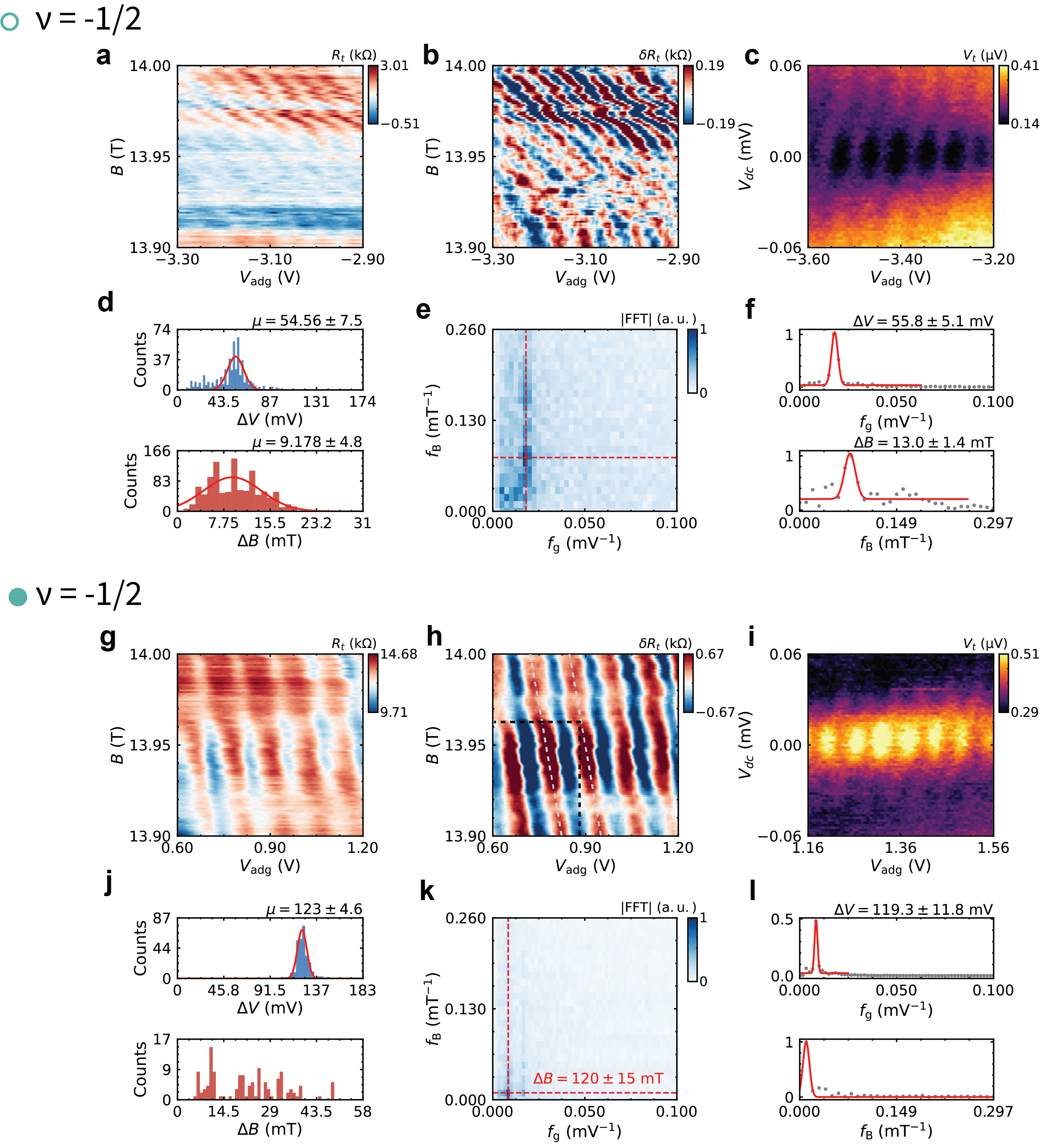}
 \begin{center}
 \caption{\textbf{Full dataset for $\nu = -1/2$.} Panels \textbf{a-f} correspond to the minimal excitation, while panels \textbf{g-l} to the double gate period regime. \textbf{a,g}, Raw $V_\mathrm{adg}$ versus $B$ map. \textbf{b,h}, The data after background subtraction. \textbf{c,i}, DC-bias dependence of oscillations, measuring $V_\mathrm{t}$ as explained in Measurement methodology. \textbf{d,j}, Histograms of the extracted peak-to-peak spacings in gate voltage, $\Delta V$, and magnetic field, $\Delta B$. \textbf{e,k}, 2D-FFT, and \textbf{f,l}, are line cuts through the 2D-FFT along the red dashed lines. For the double regime, a clear peak cannot be extracted and fitted for the magnetic field period in \textbf{k,l}, since there is no full magnetic field period present in the data of \textbf{g,h}. Here, to extract the magnetic field period we estimate the half period graphically as shown in \textbf{h}, and from this we estimate the full period $\Delta B$, which is shown in the red text in \textbf{k}.}
 \label{fig:sup_m12}
 \end{center}
\end{figure*}

\begin{figure*}[tph!]
 \renewcommand{\thefigure}{S10}
 \includegraphics[width = 0.8\textwidth]{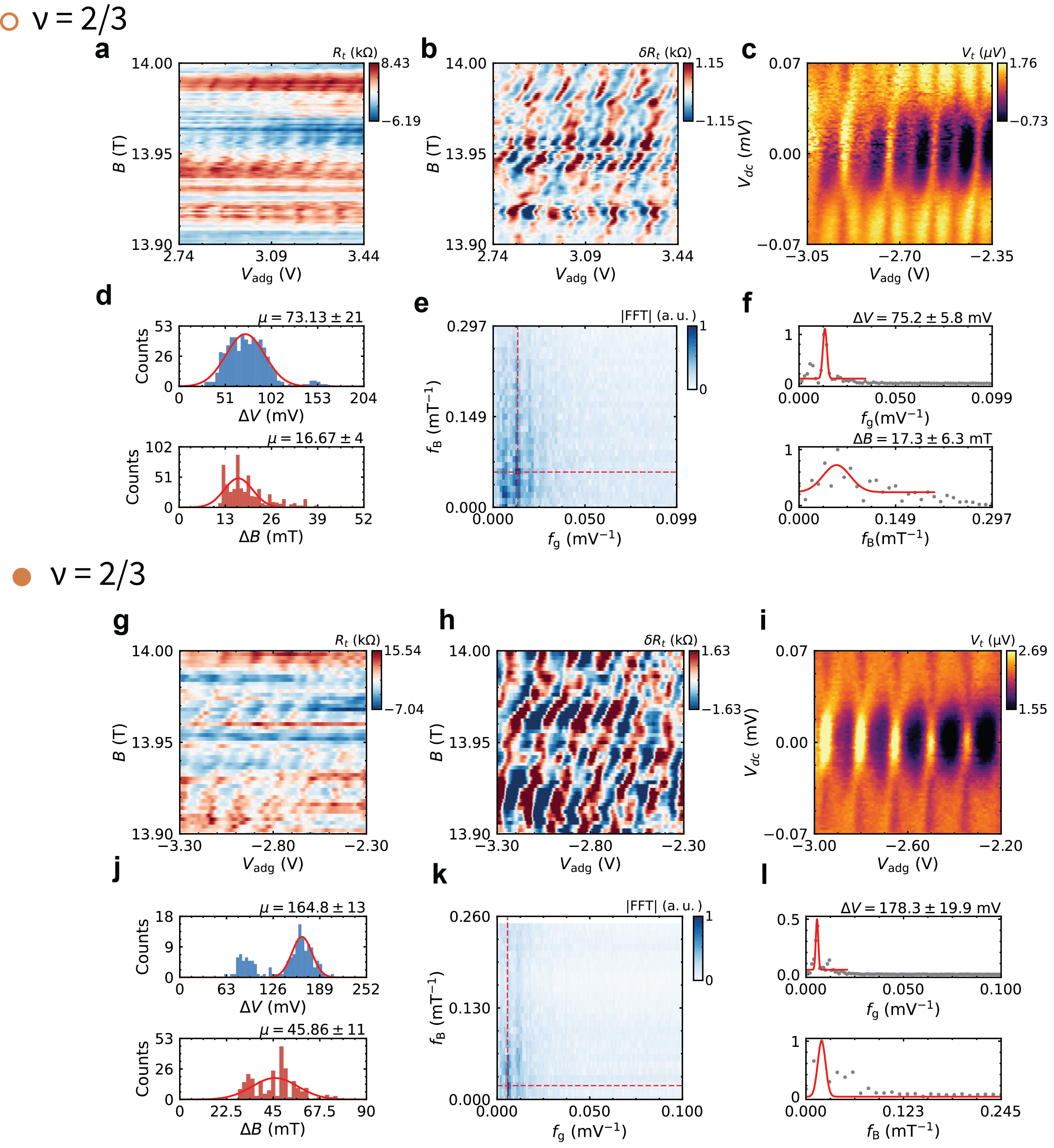}
 \begin{center}
 \caption{\textbf{Full dataset for $\nu = 2/3$.} Panels \textbf{a-f} correspond to the minimal excitation, while panels \textbf{g-l} to the double gate period regime. \textbf{a,g}, Raw $V_\mathrm{adg}$ versus $B$ map. \textbf{b,h}, The data after background subtraction. \textbf{c,i}, DC-bias dependence of oscillations, measuring $V_\mathrm{t}$ as explained in Measurement methodology. \textbf{d,j}, Histograms of the extracted peak-to-peak spacings in gate voltage, $\Delta V$, and magnetic field, $\Delta B$. \textbf{e,k}, 2D-FFT, and \textbf{f,l}, are line cuts through the 2D-FFT along the red dashed lines. For the doubling regime, the magnetic-field period extracted from the 2D-FFT line cut is not considered reliable. Inspection of the maps in \textbf{g,h} reveals multiple periodicities, which we attribute to the instability present during the measurement, as discussed in Supplementary Section~\ref{sec:histograms}. We therefore use the magnetic-field periodicity extracted from the histogram for the double regime, shown in Fig.~\ref{fig:histograms_2_3_2e_3}, where a larger error is included to reflect on this effect. For the minimal excitation magnetic field period, the histogram method and 2D-FFT analysis agree so we chose to use the 2D-FFT results for consistency.}
 \label{fig:sup_m23}
 \end{center}
\end{figure*}

\begin{figure*}[tph!]
 \renewcommand{\thefigure}{S11}
 \includegraphics[width = 0.8\textwidth]{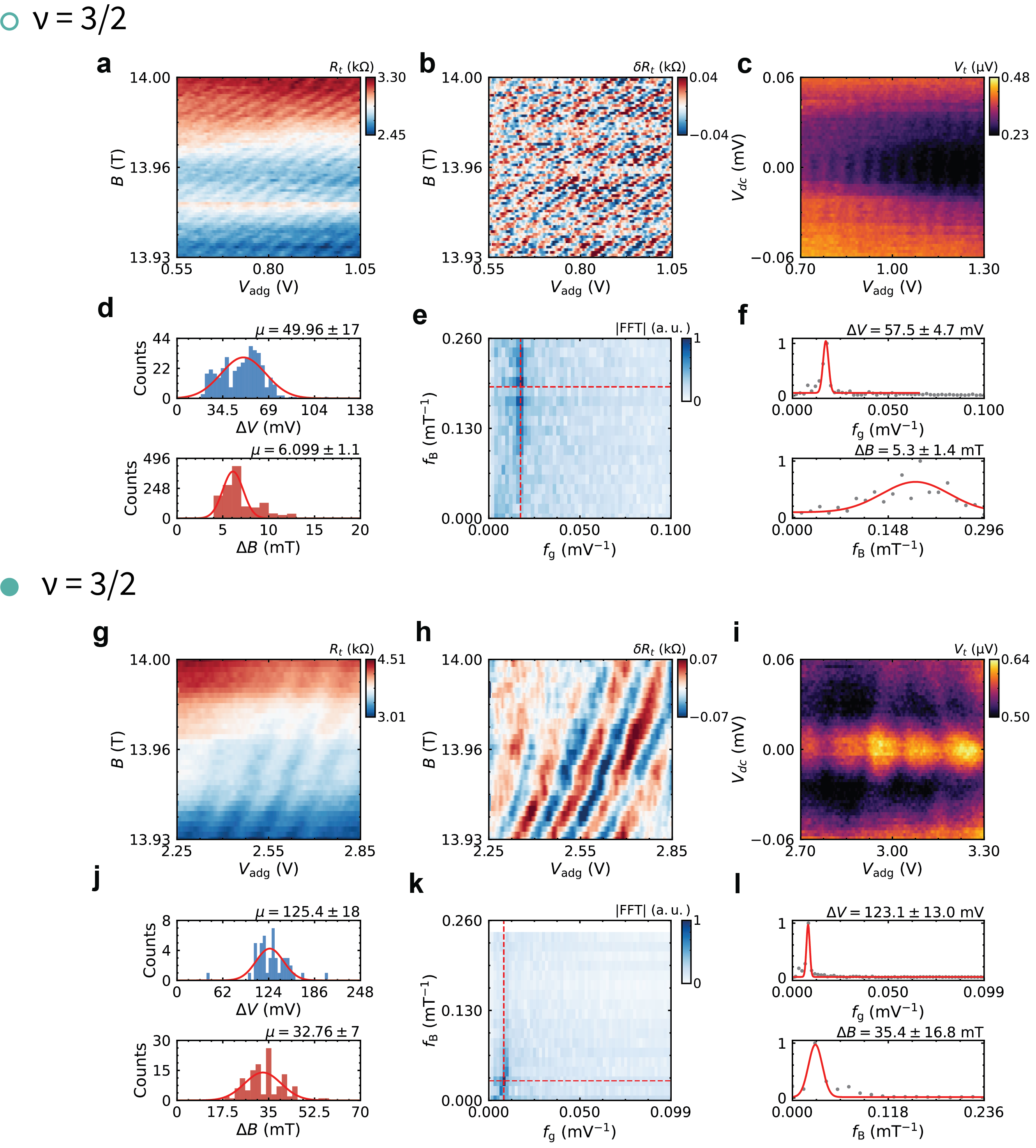}
 \begin{center}
 \caption{\textbf{Full dataset for $\nu = 3/2$.} Panels \textbf{a-f} correspond to the minimal excitation, while panels \textbf{g-l} to the double gate period regime. \textbf{a,g}, Raw $V_\mathrm{adg}$ versus $B$ map. \textbf{b,h}, The data after background subtraction. \textbf{c,i}, DC-bias dependence of oscillations, measuring $V_\mathrm{t}$ as explained in Measurement methodology. \textbf{d,j}, Histograms of the extracted peak-to-peak spacings in gate voltage, $\Delta V$, and magnetic field, $\Delta B$. \textbf{e,k}, 2D-FFT, and \textbf{f,l}, are line cuts through the 2D-FFT along the red dashed lines.}
 \label{fig:sup_32}
 \end{center}
\end{figure*}

\begin{figure*}[tph!]
 \renewcommand{\thefigure}{S12}
 \includegraphics[width = 0.8\textwidth]{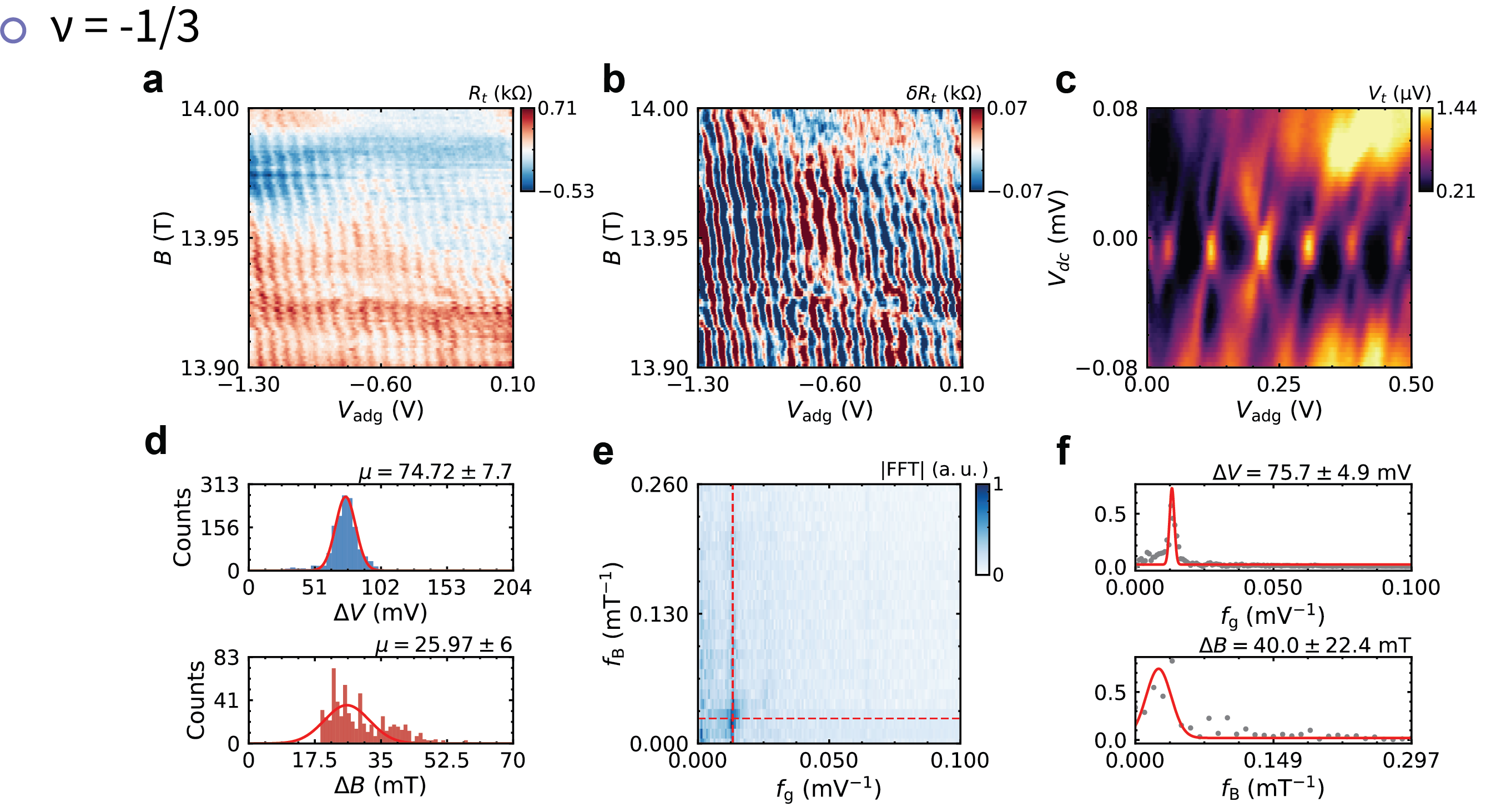}
 \begin{center}
 \caption{\textbf{Full dataset for $\nu = -1/3$.} \textbf{a}, Raw $V_\mathrm{adg}$ versus $B$ map. \textbf{b}, The data after background subtraction. \textbf{c},  DC-bias dependence of oscillations, measuring $V_\mathrm{t}$ as explained in Measurement methodology. \textbf{d}, Histograms of the extracted peak-to-peak spacings in gate voltage, $\Delta V$, and magnetic field, $\Delta B$. \textbf{e}, 2D-FFT, and \textbf{f}, are line cuts through the 2D-FFT along the red dashed lines.}
 \label{fig:sup_m13}
 \end{center}
\end{figure*}

\clearpage

\subsection{Histogram analysis at $\nu = 2/3$}
\label{sec:histograms}

\begin{figure*}[tph!]
 \renewcommand{\thefigure}{S13}
 \includegraphics[width = 0.7\textwidth]{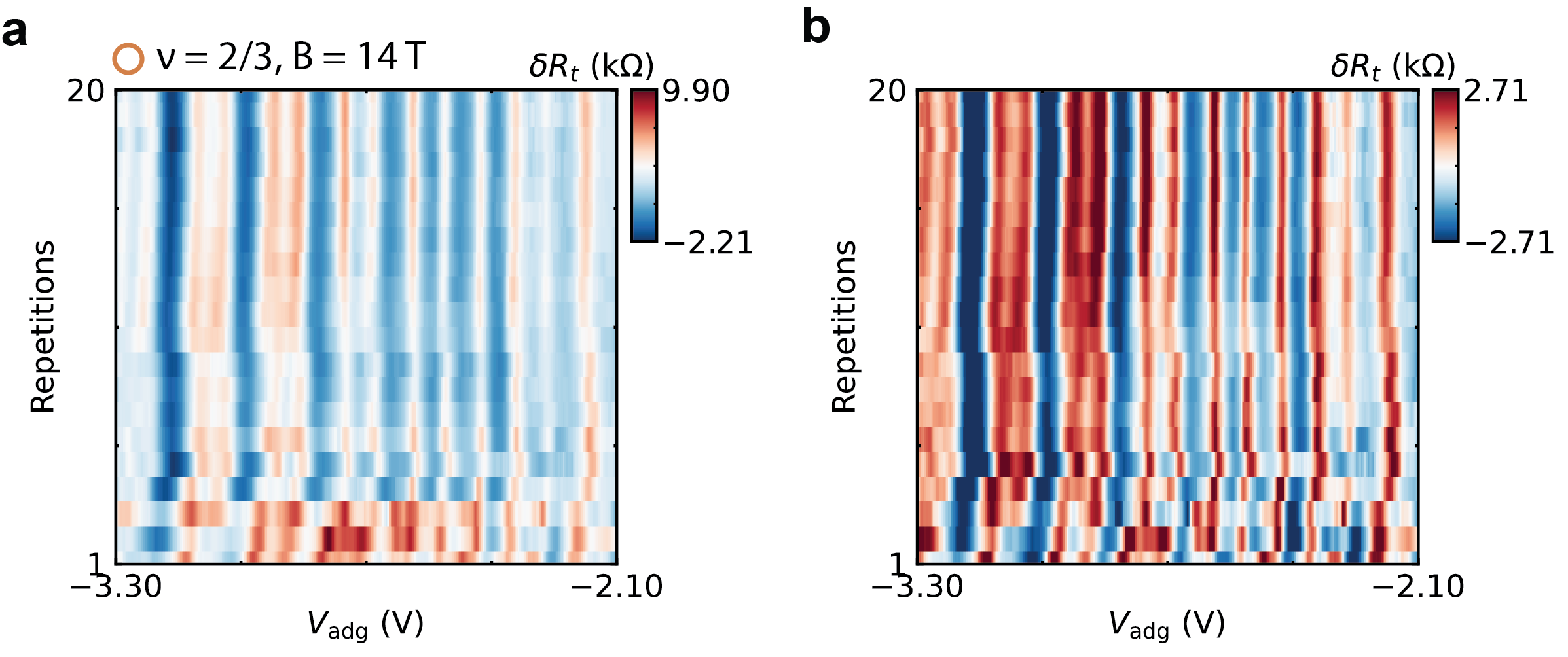}
 \begin{center}
 \caption{\textbf{Instability at $\nu=2/3$.} Data corresponding to the minimal-excitation $e/3$ oscillations at $\nu=2/3$ taken at $B = 14$\,T. \textbf{a}, Raw data from 20 consecutive repetitions of the same measurement, at a fixed magnetic field of $B = 14$\,T. The first 6 repetitions show substantial shifts between successive traces, consistent with significant gate hysteresis at this filling factor. \textbf{b}, Same data after subtraction of a third-order polynomial background.}
 \label{fig:hysteresis}
 \end{center}
\end{figure*}

For filling factor $\nu=2/3$, visual inspection of the $V_\mathrm{adg}$ versus $B$ maps indicates some instability in the oscillation pattern. As a result, the gate-voltage and magnetic-field periodicities extracted from the 2D-FFT cannot directly be considered reliable. To understand the origin of this behavior, we measured the oscillations as a function of ADG voltage at fixed magnetic field and repeated the same sweep multiple times. As shown in Fig.~\ref{fig:hysteresis}, the first several repetitions exhibit a pronounced gate-hysteresis effect before converging toward a more reproducible response. Consequently, when constructing a $V_\mathrm{adg}$ versus $B$ map by stepping the magnetic field and recording one gate-voltage trace at each field value, each horizontal trace effectively samples the system before it has reached this stationary response. This history dependence obscures the systematic evolution of the oscillations with magnetic field and therefore weakens the well-defined slope expected in the two-dimensional map, making the corresponding extracted periodicities from the 2D-FFT unreliable. So for $\nu = 2/3$ we adopt a different approach, analogous to the one used in interferometry measurements in the presence of phase jumps~\cite{Werkmeister2024Aug}. In our case the instability cannot be related to quasiparticles entering and leaving the AD-bulk. We attribute it instead to the coexistence of two configurations of comparable energy, favouring the localisation of either $e/3$ or $2e/3$ quasiparticles on the AD. A small change in the magnetic field, and therefore in the filling factor, or in the coupling to the side gates is then sufficient to switch the system between the two, substantially modifying the oscillation pattern.

\begin{figure*}[tph!]
 \renewcommand{\thefigure}{S14}
 \includegraphics[width = 0.95\textwidth]{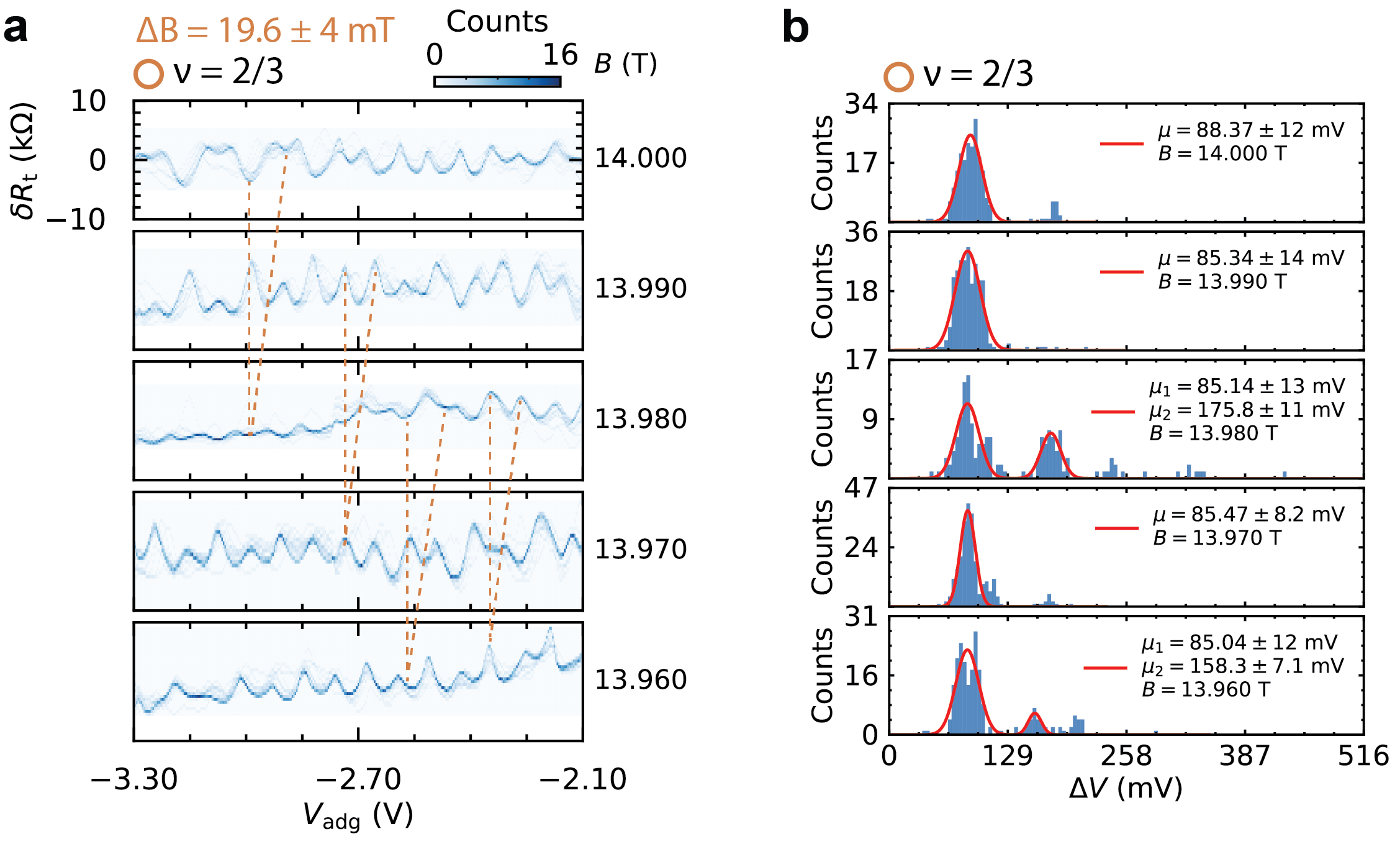}
 \begin{center}
\caption{\textbf{Histogram analysis of $\nu = 2/3$ oscillations, minimal excitation.} \textbf{a}, Two-dimensional histograms of the repeated antidot-gate sweeps acquired at fixed magnetic field values. At each $V_\mathrm{adg}$, the measured resistance values are binned over all repetitions, with the color scale indicating the number of counts in each resistance bin. The panels are shown for different values of $B$. \textbf{b}, Distribution of the gate-voltage spacings $\Delta V$ extracted from the oscillations at each magnetic field. Consecutive peak-to-peak or dip-to-dip spacings are collected over all repetitions and fitted with one or two Gaussians, depending on whether one or two characteristic periodicities are resolved. The fitted Gaussian centers give the corresponding gate-voltage periodicities.}
 \label{fig:histograms_2_3_e_3}
 \end{center}
\end{figure*}

\begin{figure*}[tph!]
 \renewcommand{\thefigure}{S15}
 \includegraphics[width = 0.95\textwidth]{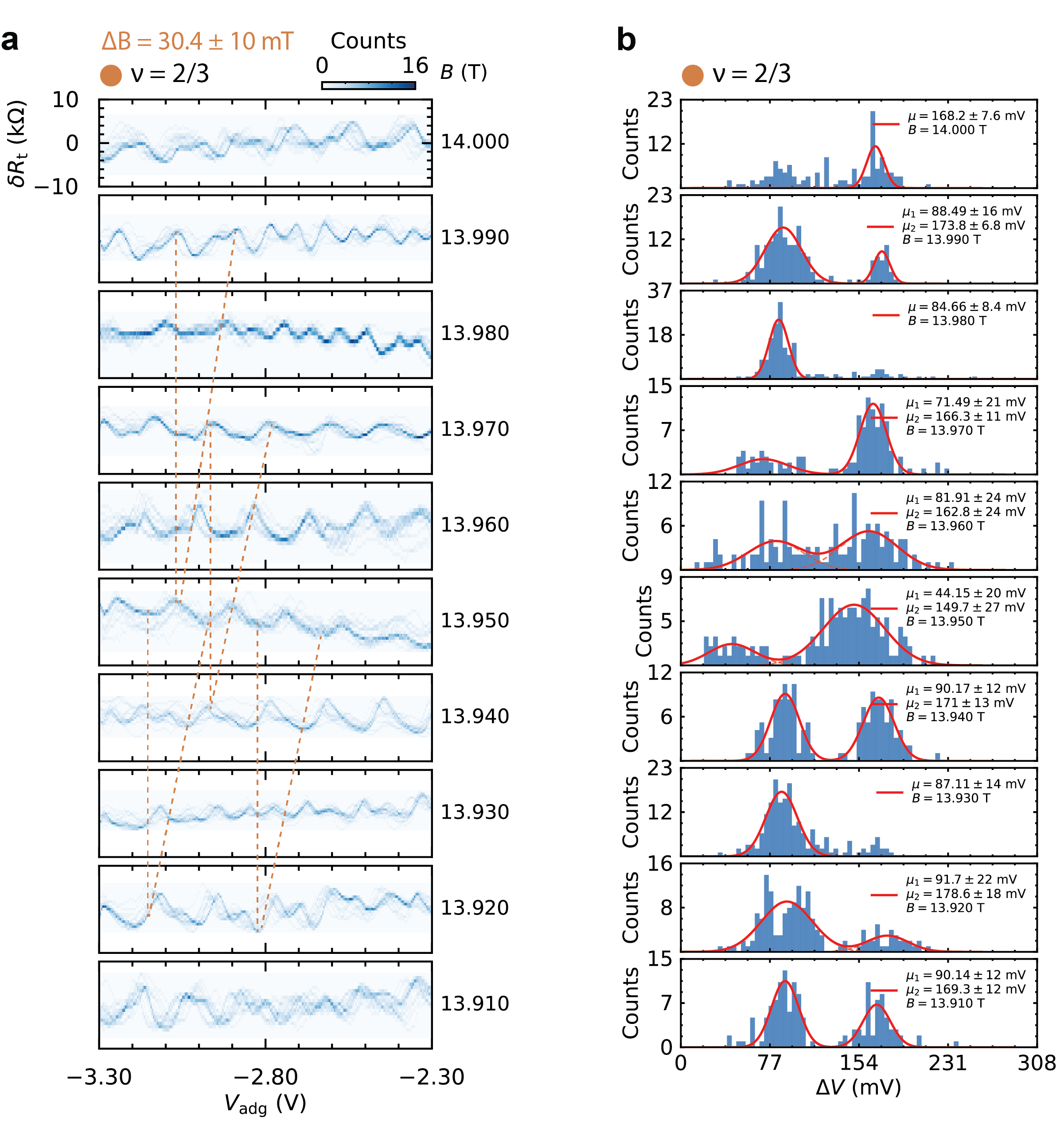}
 \begin{center}
\caption{\textbf{Histogram analysis of $\nu = 2/3$ oscillations, doubling regime.} \textbf{a}, Two-dimensional histograms of the repeated antidot-gate sweeps acquired at fixed magnetic field values. At each $V_\mathrm{adg}$, the measured resistance values are binned over all repetitions, with the color scale indicating the number of counts in each resistance bin. The panels are shown for different values of $B$. \textbf{b}, Distribution of the gate-voltage spacings $\Delta V$ extracted from the oscillations at each magnetic field. Consecutive peak-to-peak or dip-to-dip spacings are collected over all repetitions and fitted with one or two Gaussians, depending on whether one or two characteristic periodicities are resolved. The fitted Gaussian centers give the corresponding gate-voltage periodicities.}
 \label{fig:histograms_2_3_2e_3}
 \end{center}
\end{figure*}

We repeat the measurement shown in Fig.~\ref{fig:hysteresis} for different magnetic field values. For each fixed magnetic field $B$, the repeated gate-voltage sweeps are first represented as a two-dimensional histogram of the measured resistance $\delta R_\mathrm{t}$ as a function of gate voltage. At each gate voltage, the resistance values obtained over all repetitions are binned, such that the color scale represents the number of occurrences within each resistance bin. This representation provides a direct visualization of the reproducibility and evolution of the oscillations over repeated sweeps. This is shown in Fig.~\ref{fig:histograms_2_3_e_3}a and \ref{fig:histograms_2_3_2e_3}a for the minimal excitation oscillations and doubling regime, respectively.

We then track a selected peak or dip vertically as a function of $B$ until one complete oscillation cycle is completed. Several examples of such trajectories are shown following the orange dashed lines in Figs.~\ref{fig:histograms_2_3_e_3}a and \ref{fig:histograms_2_3_2e_3}a. Repeating this procedure for several peaks and dips yields a set of full cycle magnetic field periods. The magnetic-field periodicity reported at the top of each figure is obtained by averaging these values, while the corresponding uncertainty is taken as the standard deviation of the extracted periods. This procedure yields a magnetic-field periodicity consistent with the value obtained from the 2D-FFT. 

For the minimal excitation oscillations, using this method, we find a magnetic field periodicity equal to $\Delta B = 19.6 \pm 4$\,mT, and for the doubling regime $\Delta B = 30.4 \pm 10$\,mT.

To quantify the gate voltage periodicity, for each magnetic field, peaks are subsequently identified independently in each trace (each line of Fig.\ref{fig:hysteresis}), and the spacings $\Delta V_{\mathrm{adg}}$ between consecutive peaks are extracted. The spacings obtained from all repetitions at the same magnetic field are pooled into a single distribution and displayed as a histogram. This distribution is fitted with a single Gaussian when one dominant periodicity is observed, or with the sum of two Gaussians when two distinct populations of spacings are resolved. When there is a hint of a second peak, but the sample size is not substantial, we could not fit a second Gaussian. The Gaussian centers, $\mu_{i}$, provide the characteristic gate-voltage periods, while the corresponding widths, $\sigma_{i}$, quantify the spread of the extracted periodicities. 

\clearpage

\subsection{Temperature dependence}

To gain further insight into the characteristic energy scales associated with the states investigated in this work, we study the temperature dependence of the AD oscillations for different filling factors. Increasing the temperature introduces thermal broadening, which progressively suppresses the oscillations when the thermal energy becomes comparable to the energy scales governing transport through the AD. 

In the following discussion, for brevity we refer to the minimal-excitation oscillations of the even-denominator states as $e/4$ oscillations and to the doubled gate-period regime as $e/2$ oscillations. Similarly, at $\nu=2/3$ we refer to the minimal-excitation and doubled-period regimes as $e/3$ and $2e/3$ oscillations, respectively. These labels are used only as shorthand for the observed gate-periodicities and do not imply that the doubled-period regimes arise from microscopic bunching or from tunneling of quasiparticles with charge $e/2$ or $2e/3$.

\begin{figure*}[tph!]
 \renewcommand{\thefigure}{S16}
 \includegraphics[width = 0.95\textwidth]{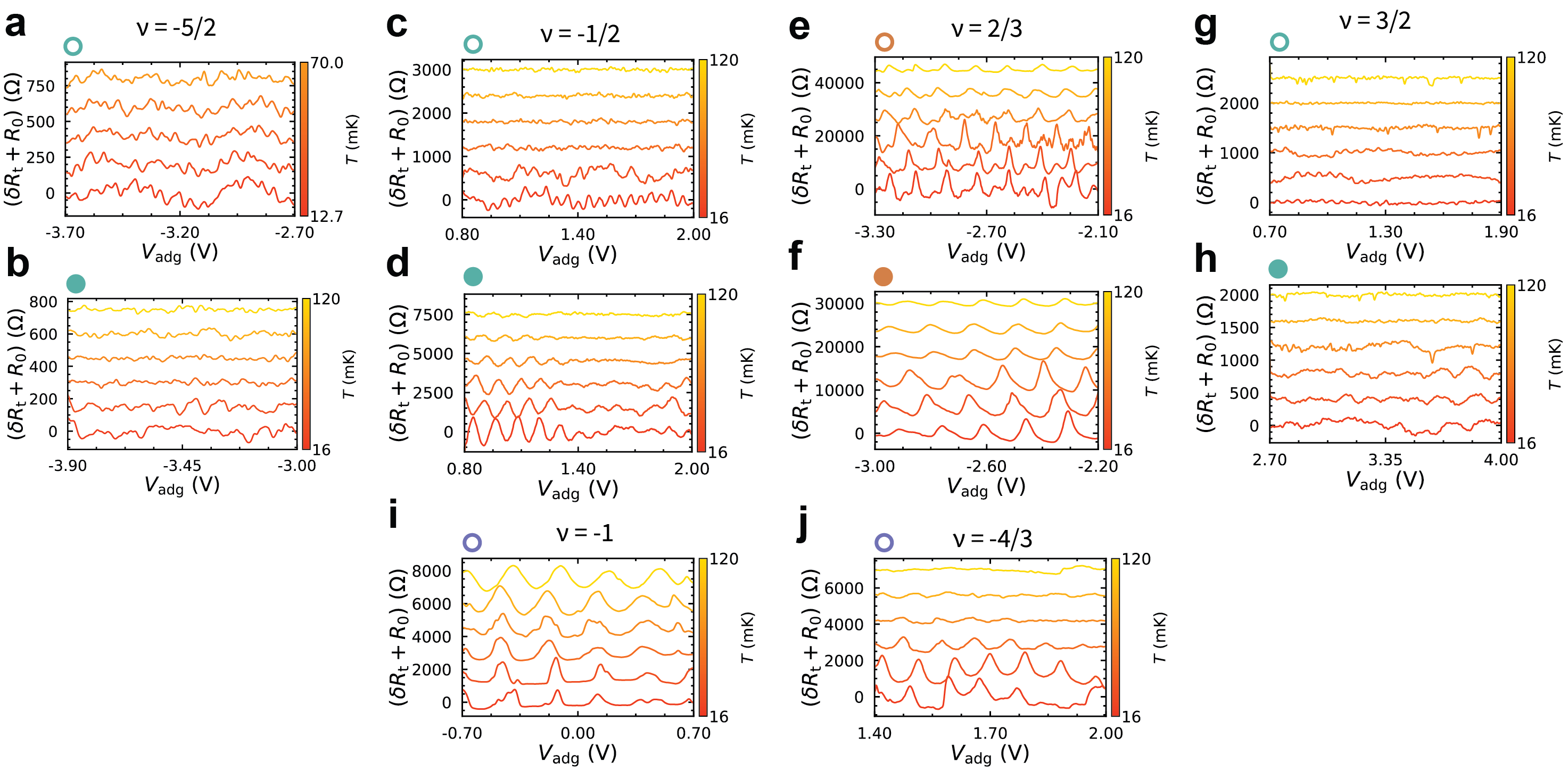}
 \begin{center}
\caption{\textbf{Temperature evolution of $\delta R_\mathrm{t}$ oscillations.} Open circles denote oscillations associated with the minimal excitation, while filled circles denote the doubling regime. Colors distinguish the different classes of quantum Hall states, following the convention of Fig.~\ref{fig:figure_4}: turquoise denotes even-denominator states, ochre hole-conjugate states, and light-purple particle-like and integer states. Starting from the coldest trace, the temperature steps are $T = 16, \ 30, \ 50, \ 70, \ 95, \ 120$\,mK for all panels except \textbf{a}, where the steps are $T = 12.7, \ 20, \ 35, \ 50, \ 70$\,mK. \textbf{a,b}, Oscillations corresponding to $e/4$ and $e/2$, respectively, at $\nu=-5/2$. \textbf{c,d}, Oscillations corresponding to $e/4$ and $e/2$, respectively, at $\nu=-1/2$. \textbf{e,f}, $e/3$ and $2e/3$, respectively, at $\nu=2/3$. \textbf{g,h}, Oscillations corresponding to $e/4$ and $e/2$, respectively, at $\nu=3/2$. \textbf{i}, $e$ at $\nu=-1$. \textbf{j}, $e/3$ at $\nu=-4/3$. For each trace, a polynomial background is subtracted, and traces at different temperatures are vertically offset by an arbitrary value $R_0$ for clarity. All measurements are performed at $B=13.95$\,T.}
 \label{fig:temp_dep}
 \end{center}
\end{figure*}

As shown in Fig.~\ref{fig:temp_dep}a,c,g, the minimal excitation $e/4$ oscillations at the even-denominator states $\nu=-5/2, -1/2$ and $3/2$, respectively, are suppressed between $50$\,mK and $70$\,mK. In the doubling regime, however, the temperature dependence differs between these states. At $\nu=-5/2$ and $3/2$, the $e/2$ oscillations (see Fig.~\ref{fig:temp_dep}b,h) are similarly suppressed above approximately $50$\,mK. In contrast, at $\nu=-1/2$, the $e/2$ oscillations (see Fig.~\ref{fig:temp_dep}d) persist to significantly higher temperatures. Their amplitude decreases with increasing temperature but remains finite up to $120$\,mK. A similar behavior is observed in the doubling regime of $\nu = 2/3$, where the double gate-period oscillations also remain finite up to $120$\,mK.

\begin{figure*}[tph!]
 \renewcommand{\thefigure}{S17}
 \includegraphics[width = 0.95\textwidth]{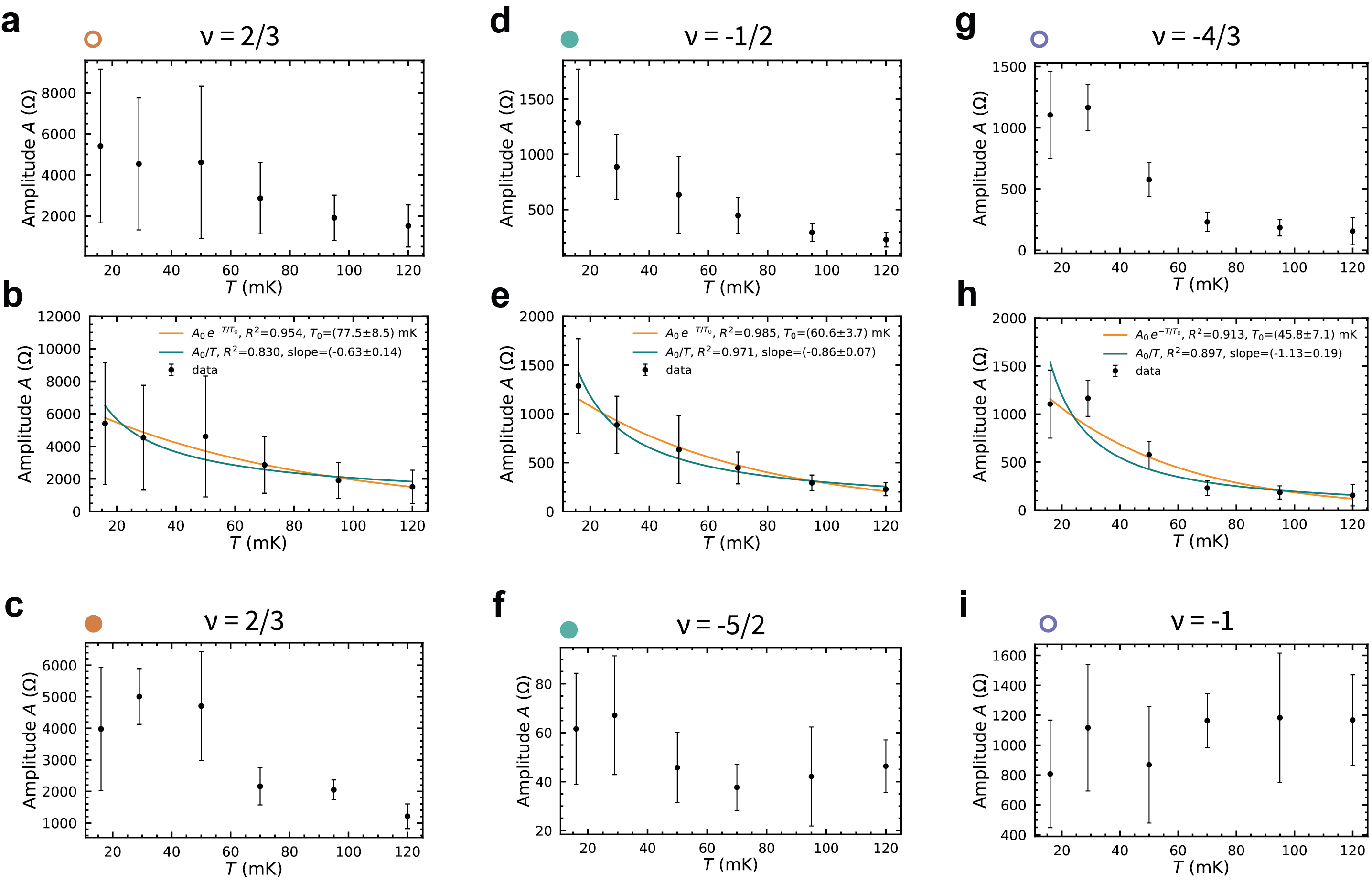}
 \begin{center}
\caption{\textbf{Temperature dependence of the oscillation amplitude $A$.} Open circles denote oscillations associated with the minimal excitation, filled circles the doubling regime. Colours distinguish the classes of quantum Hall states, following the convention of Fig.~\ref{fig:figure_4}: turquoise for even-denominator states, ochre for hole-conjugate states and light purple for particle-like and integer states. \textbf{a--i} Amplitude $A$ as a function of temperature $T$ for the $e/3$ oscillations at $\nu = 2/3$ (\textbf{a}, fit in \textbf{b}), the $2e/3$ oscillations at $\nu = 2/3$ (\textbf{c}), the $e/4$ oscillations at $\nu = -1/2$ (\textbf{d}, fit in \textbf{e}), the oscillations at $\nu = -5/2$ (\textbf{f}), at $\nu = -4/3$ (\textbf{g}, fit in \textbf{h}) and at $\nu = -1$ (\textbf{i}). In \textbf{b}, \textbf{e} and \textbf{h} the orange line is an exponential fit (Eq.~\eqref{eq:exp_model}) and the blue line a power-law fit (Eq.~\eqref{eq:pow_model}).}
 \label{fig:temp_dep_fit}
 \end{center}
\end{figure*}

For the oscillations where we could accurately extract the amplitude A, we fit the oscillation amplitude versus temperature with two functional forms, corresponding to two distinct physical origins of the oscillations~\cite{Sim2008Feb, Moreau2022Mar} in Fig.~\ref{fig:temp_dep_fit}.

For an Aharonov-Bohm origin, the amplitude is expected to decay exponentially with temperature, reflecting the thermal dephasing of the interfering
paths~\cite{Martins2013Mar}. In this case we fit

\begin{equation}
 A(T) = A_0 \, e^{-T/T_0},
 \label{eq:exp_model}
\end{equation}

\noindent with $A_0$ the amplitude extrapolated to $T = 0$ and $T_0$ the thermal damping scale governing the decay of the AB visibility in an interferometer~\cite{Neder2007Jul},

\begin{equation}
 k_B T_0 = \frac{\hbar v}{4\pi^2 R}.
 \label{eq:thermal_scale}
\end{equation}

This scale differs only by a numerical factor from the single-level energy splitting of the AD-bound states in the absence of electron-electron interactions,

\begin{equation}
 \Delta E \simeq \frac{2\hbar v}{D},
 \label{eq:ch1_energy_splitting}
\end{equation}

\noindent so that $\Delta E = 4\pi^2 k_B T_0$. The two quantities are set by the round-trip time $\tau = 2\pi R / v$ of a quasiparticle around the loop.

For a Coulomb-blockade origin, the amplitude is instead expected to follow a power law~\cite{Glazman1989Aug, Karakurt2001Sep, Hackens2010Jul},

\begin{equation}
 A(T) = A_0 \, T^{-\alpha},
 \label{eq:pow_model}
\end{equation}

\noindent where $\alpha \approx 1$ corresponds to the $T^{-1}$ dependence expected in the CD regime. 

The two models are compared through the coefficient of determination $R^2$ of their respective linearised regressions.

Figures~\ref{fig:temp_dep_fit}b,e,h show both fits for $\nu = 2/3$, $\nu = -1/2$ and $\nu = -4/3$. For several of the other oscillations, including the minimal excitation oscillations of the even-denominator states, the amplitude is suppressed too rapidly with temperature for a reliable extraction.

For the remaining filling factors (Fig.~\ref{fig:temp_dep_fit}c,f,i) neither model describes the data satisfactorily, as reflected by the low $R^2$ values, and the corresponding fits are therefore not shown. Although the fits cannot therefore identify the transport regime, they still constrain the electron temperature of our setup. All the $e/4$ oscillations, which have comparable energy scales, are suppressed in the same range of \SIrange{30}{50}{\milli\kelvin}, which places an upper bound on the electron temperature. 

We stress that the present data extend only to $T \approx 2\,T_0$, a range over which the two models are not strongly distinguishable: below $T_0$ both are slowly varying and reproduce the data with comparable quality, despite describing fundamentally different mechanisms. This is what we observe for the $e/4$ oscillations at $\nu = -1/2$ and the $e/3$ oscillations at $\nu = 2/3$, which are described equally well by Eq.~\eqref{eq:exp_model} and Eq.~\eqref{eq:pow_model} (Fig.~\ref{fig:temp_dep_fit}b,e).

\clearpage

\subsection{Expected periodicity}

\begin{figure*}[tph!]
 \renewcommand{\thefigure}{S18}
 \includegraphics[width = 0.5\textwidth]{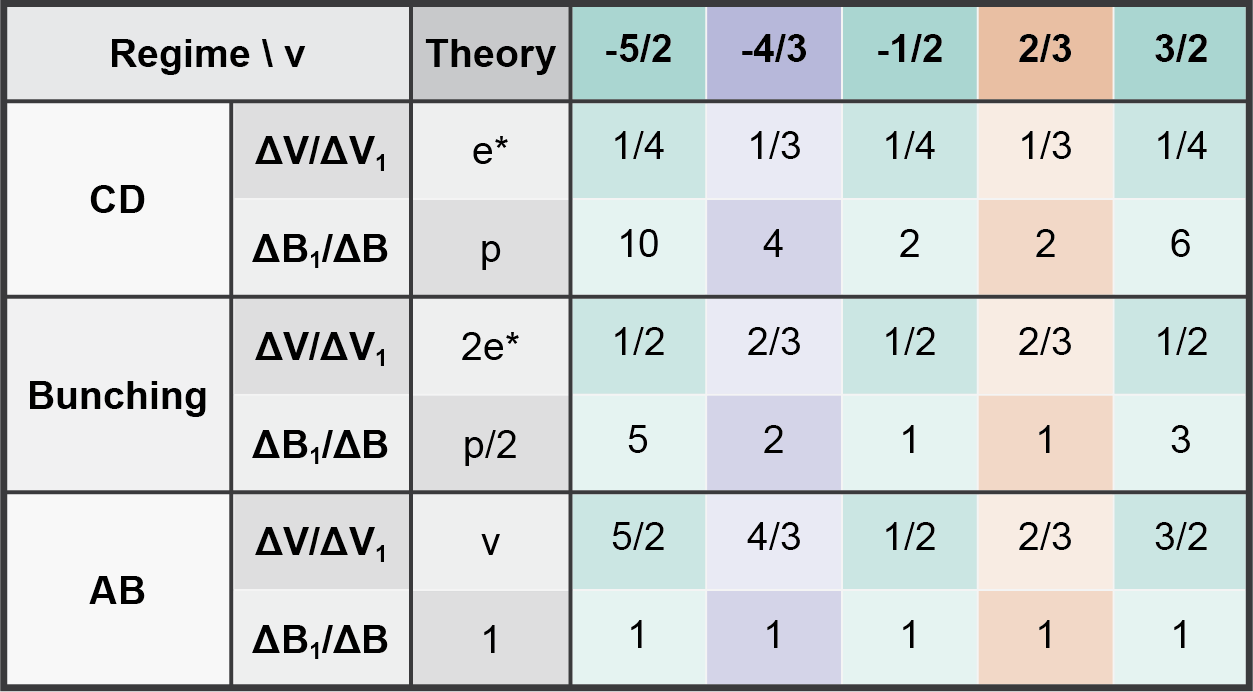}
 \begin{center}
\caption{\textbf{Table resuming the possible scenarios and the respective periodicities} Table resuming the possible scenarios and the respective periodicities. We have reported the expected period of $\nu = -5/2, -4/3, -1/2, 2/3$ and $3/2$. The theory refers to Ref.~\cite{LevySchreier2016Aug}.}
 \label{fig:table}
 \end{center}
\end{figure*}

\end{document}